\documentclass[11pt]{article}
\usepackage{a4wide}
\usepackage[T1]{fontenc}
\usepackage{graphicx}
\usepackage[misc,geometry]{ifsym}

\usepackage{amsmath,amssymb, amsfonts, dsfont, bbm,calc,mathtools,commath,url}
\usepackage{enumerate}
\usepackage{soul}
\usepackage{nicefrac}
\usepackage{bussproofs}
\usepackage{multirow}
\usepackage{lipsum}
\usepackage{synttree}
\usepackage{xcolor}
\usepackage{multicol}
\usepackage{comment}

\usepackage[dvipsnames]{xcolor}

\usepackage{enumerate}
\usepackage{enumitem}
\setitemize{noitemsep,parsep=0pt,partopsep=0pt}

\def\phi{\varphi}

\newcommand{\mathrmL}{{\mathchoice{\mbox{\rm\L}}{\mbox{\rm\L}}{\mbox{\rm\scriptsize\L}}{\mbox{\rm\tiny\L}}}}

\newtheorem{theorem}{Theorem}[section]
\newtheorem{definition}[theorem]{Definition}
\newtheorem{lemma}[theorem]{Lemma}
\newtheorem{corollary}[theorem]{Corollary}
\newtheorem{observation}[theorem]{Observation}

\newtheorem{example}[theorem]{Example}

\newcommand{\qed}{{\hspace*{\fill} {$\cal QED$}}}
\newenvironment{proof}
{\begin{trivlist} \item[] {\it Proof.}}{\qed \end{trivlist}}

\newcommand{\lang}[1]{\ensuremath{\mathcal #1}}
\newcommand{\Fm}{\ensuremath{\mathit{Fm}_{\mathrmL}}}
\newcommand{\Tm}{\Fm}

\newcommand{\Vbl}{\ensuremath{\mathrm{Var}}}

\newcommand{\logic}[1]{\ensuremath{\mathrm {#1}}}

\newcommand\Luk{\logic{\mathrmL}}

\newcommand{\alg}[1]{ {\ensuremath{\mathcal {#1}}}}

\newcommand\SAT{\ensuremath{\mathbf{SAT}}}

\newcommand\standardL{\ensuremath{{[0,1]_\mathrmL}}}

\newcommand\finiteMV[1]{ \ensuremath{ \mathrm{MV}_{#1} }}
\newcommand\MVN{\ensuremath{\finiteMV{N}}}

\newcommand\ComplexityClass[1]{\ensuremath{\mathbf{#1}}}
\newcommand\Po{\ensuremath{\ComplexityClass{P}}}
\newcommand\NP{\ensuremath{\ComplexityClass{NP}}}

\newcommand\FPNP{\ensuremath{\ComplexityClass{FP}^{\ComplexityClass{NP}}}}

\begin{document}

\title{Algorithmic properties of the MaxSAT problem in MV-algebras}

\author{Zuzana Hanikov\'a$^1$, Felip Many{\`{a}}$^{2,}\footnote{Corresponding author.}$, Amanda Vidal$^1$\\
$^1$ Institute of Computer Science of the Czech Academy of Sciences\\
{ \tt  \{amanda,zuzana\}@cs.cas.cz }\\
$^2$ Artificial Intelligence Research Institute of the CSIC\\
{ \tt  felip@iiia.csic.es }
}

\maketitle             

\begin{abstract}
This work investigates the computational properties of the maximum satisfiability (MaxSAT) problem for multisets of arbitrary formulas of propositional Łukasiewicz logic, with truth values in the real interval $[0,1]$ or in one of its finite subchains.  The decision version of infinite-valued MaxSAT is polynomial-time reducible to satisfiability in the same logic, hence in NP. The computational complexity of the corresponding optimization problem is studied using the framework provided by Krentel. Fragments of the language are also explored, with an emphasis on the notion of clauses, and some results familiar from the Boolean semantics are reproduced. The MaxSAT problem for fixed finite MV-chains is reduced to  Boolean satisfiability and partial maximum satisfiability, along with a discussion of the merit of these results for the infinite-valued analogues. Finally,  analytic methods of addressing satisfiability and maximum satisfiability is provided,
following the approaches of Hähnle and Olivetti, and soundness and completeness is proved.
\end{abstract}

\section{Introduction}

Owing to its generality, satisfiability of propositional formulas in the two-valued Boolean semantics is arguably the most prominent computational problem currently under consideration. To formulate this problem, it is enough to consider a (denumerable) set of propositional atoms, each of which can take the values `true' and `false', and combine them using the usual connectives such as conjunction, disjunction, and negation, obtaining \emph{propositional formulas}, which for our ends and purposes can be identified with \emph{Boolean terms}. A propositional formula is satisfiable provided there is a Boolean assignment under which the formula is true. The \emph{satisfiability problem}, informally speaking, is the task to decide which formulas are satisfiable in the Boolean semantics; formally, the problem is just the set of positive instances. Quite often the satisfiability problem is considered for instances in \emph{conjunctive normal form:} each instance is a multiset of clauses, where each clause is a disjunction of literals. Below in Section \ref{s_easy_hard} of this paper, this formulation of the satisfiability problem will be referred to as {\bf Boolean-SAT}. The Boolean satisfiability problem for propositional formulas in conjunctive normal form was the first problem to have been proved complete for the class $\NP$ under polynomial-time reductions   \cite{Cook:Complexity1971}. It continues to be a computational benchmark, both in a theoretical and in a practical sense.

On this view, satisfiability is a \emph{semantic} problem: it relates not only to a logic, but primarily to a semantics for the language of the logic.
In this work, we will be interested in propositional formulas of the language of {\L}ukasiewicz logic, 
and in their semantics provided by the MV-algebra on the real unit interval with the natural order,  often referred to as the ``standard MV-algebra'' and denoted $\standardL$. Next to that, we will also consider its finite subalgebras. Within this paper, the $\SAT$ problem is the task to determine, for an arbitrary given formula $\phi$ of the language of {\L}ukasiewicz logic, whether or not it has a satisfying assignment in the algebra $\standardL$, where an assignment satisfies a formula if it sends it to the value $1$ (the top element of an MV-algebra, its only distinguished element, in analogy to the Boolean case). The notation---which omits any qualifications---indicates that the semantics of the standard MV-algebra is  prominent in this work;
if another algebra, distinct from $\standardL$, is considered as semantics for satisfiability problems, it will be indicated by explicit change in notation.
The language of {\L}ukasiewicz logic is richer than that of Boolean logic (for example,
it features the ``multiplicative'' conjunction $\odot$ as well as the ``additive'' conjunction $\land$). This is in keeping with the fact that Boolean logic
is formally stronger than {\L}ukasiewicz logic, and in the former, some pairs of connectives may coincide. Accordingly, Boolean algebras are (termwise equivalent to) a subclass of MV-algebras and the satisfiability problems considered here for the MV-case are indeed a generalization of the Boolean case.

Given that the algebra $\standardL$ has an infinite domain, it does not follow that satisfiability is a decidable problem.\footnote{If one varies the choice of an MV-algebra, one obtains various sets of satisfiable formulas. 
It is not hard to show that there are continuum many pairwise distinct sets of formulas of the language of {\L}ukasiewicz logic, each of which is the set of satisfiable formulas of some subalgebra of $\standardL$ (see \cite[Theorem 6.6]{Hanikova-Savicky:SAT}). As a consequence, most of these sets are undecidable.}
A seminal result of Mundici \cite{Mundici:Satisfiability} established that the satisfiability problem is decidable and actually $\NP$-complete,
 where the containment in $\NP$ is obtained by providing a polynomial-size upper bound to the common denominator for at least one satisfying assignment. 
 It follows that the formula is satisfied in a finite totally ordered MV-algebra (also known as finite MV-chain) with a suitable denominator. 
This line of research was continued in \cite{Aguzzoli-Ciabattoni:Finiteness,Aguzzoli:asymptoticBoundLuk,Aguzzoli-Gerla:finitereductions,Aguzzoli-Gerla-Hanikova:complexityIssues},
see also \cite{SchockaertJanssenVermeirDeCock:FiniteInfiniteSAT,Hanikova:Handbook}.

This paper is primarily not about satisfiability, but rather  \emph{maximum satisfiability}, an optimization problem with a natural decision version. In the context of Boolean logic, the maximum satisfiability problem arises naturally from the satisfiability problem for conjunctive normal form: supposing that an instance (i.e., a list of clauses) is unsatisfiable,  how many of the clauses in the list can be satisfied simultaneously? It has been recognized early on by Mundici \cite{Mundici:UlamGameMaxSATandMVPart} that formulas of {\L}ukasiewicz logic are a suitable device for \emph{counting}; his paper gives a reduction of the (decision version of) the Boolean MaxSAT problem to the $\SAT$ problem; see also \cite{PretoManyaFinger:linkingLukMaxSat}. The formulation of the maximum satisfiability problem in the setting of MV-algebras follows the suit of the Boolean case, but drops any reference to normal form: given as an instance a multiset $M$ of arbitrary formulas of the language of {\L}ukasiewicz logic, the task is to find the greatest natural number $k$, over all assignments to variables in $M$, such that at least $k$ formulas can be satisfied by a single assignment.  The maximum satisfiability problem for a list of arbitrary formulas over the three-element MV-algebra (i.e., in the setting of a finite-valued semantics) was addressed in \cite{LiManyaVidal:TableauxMaxSATLuk}, using semantic tableaux; the approach therein can be generalized to other finite MV-chains, but not to MV-chains with infinitely many elements.

Our approach to the maximum satisfiability problem is primarily theoretical. We try to provide a complexity classification to assess its difficulty, for further consideration. However one ought to  bear in mind also the practical aspects of addressing the family of satisfiability and maximum satisfiability problems. Needless to say, such solvers are available for the Boolean satisfiability and Boolean maximum satisfiability problems \cite{bacchus2021maxsat,LiManya:2021maxsat}. An important method of addressing the satisfiability problem in $\standardL$ was provided by H\"ahnle, who reduced it to the Mixed Integer Programming (MIP) problem in \cite{Hahnle:MIP}. This method paves a direct route to practice, owing to the available robust, heavily optimized MIP solvers. See the discussion under Corollary \ref{cor_maxsat_npc} below for references to further works, also providing reductions to SMT-solvers.
It seems fair to say that providing a theoretical classification addresses the asymptotic complexity in the worst-case scenarios, whereas reducing to an efficient solver addresses also the average case scenarios for real-life instances.

This work is an extended version of a previous conference paper \cite{Hanikova-Manya-Vidal:MaxSAT2023} of the same authors.
 We review the differences in research content. Section \ref{s_easy_hard} is new: relying on the results in \cite[Section 7]{Hanikova-Savicky:SAT}, 
 and re-using \cite[Lemma 2]{Hanikova-Manya-Vidal:MaxSAT2023} with a simplified proof (see Lemma 3.3 \label{lm:convex_formula} below), we show that the infinite-valued {\L}ukasiewicz semantics of the algebra $\standardL$ mimics the situation familiar from the Boolean semantics, providing classes of formulas for which satisfiability is decidable in polynomial time whereas  maximum satisfiability  is $\NP$-hard. In Section \ref{s_canonical}, we have simplified the proof of Theorem \ref{th_complete_fpnp}, using the material of the previous section. (The main statement of Section \ref{s_canonical}, namely Theorem \ref{thm_maxsat_sat}, was already presented  in \cite{Hanikova-Manya-Vidal:MaxSAT2023}.) Section \ref{sec:redToClassic} is new and  addresses the finite-valued semantics for a fixed cardinality.  Section \ref{s_tableau} is without substantial changes.

Since the publication of the conference paper \cite{Hanikova-Manya-Vidal:MaxSAT2023}, the weighted version of the MaxSAT problem has also been addressed, and shown to be complete for $\FPNP$, in \cite{LapentaNapolitano:MaxSATLuk}.

We outline below the  paper structure and results.
Section \ref{s_formulation} introduces the language and semantics of {\L}ukasiewicz logic, fixes our terminology, gives definitions of the main computational problems addressed, and surveys some existing methods and target problems for reductions. Section \ref{s_easy_hard} defines the Boolean satisfiability and maximum satisfiability problems as used in this paper, provides useful technical results, namely  Lemmma \ref{lm:convex_formula}  with a simplified proof and Theorem \ref{maxsat_for_fragment_is_boolean} which shows that for multisets of clauses, the maximum satisfiability problem in the Boolean semantics coincides with that in the semantics given by $\standardL$. Using this, it is shown that for multisets of clauses with 1 or 2 literals per clause, satisfiability is tractable whereas maximum satisfiability is $\NP$-hard for the semantics $\standardL$, in parallel to the Boolean case. The end of the section discusses various language fragments from the point of view of properties of definable functions and functional completeness in the Boolean semantics. Section \ref{s_canonical} shows that the decision version of MaxSAT in $\standardL$ is in $\NP$ by reducing it to the satisfiability problem in the same semantics, in Theorem \ref{thm_maxsat_sat}. Further in the section we outline how to solve maximum satisfiability as an optimization problem in $\standardL$ with the use of the decision version as an oracle, where the number of oracle queries is logarithmic in the input size. This may be viewed as an upper bound on the complexity of the optimization problem. To complement this, we also show the problem is complete for $\FPNP[O(\log m)]$ under metric reductions, which can be taken as a lower bound, under a suitable notion of reduction provided in \cite{Krentel:ComplexityOptimization}.  
Section  \ref{sec:redToClassic} addresses the finite-valued semantics, for a fixed denominator $N$ of a simple MV-chain.
We provide a polynomial-time reduction of the decision version {\bf MaxSAT-DEC}$_N$ to Boolean satisfiability in Lemma \ref{lem:DEC-VALMAXSATtoCLASSAT} and the optimization version {\bf MaxSAT-OPT}$_N$ to the Boolean partial maximum satisfiability in Lemma \ref{lem:VALMAXSATtoCLASPARTMAXSAT}. As discussed at the end of the section, it is not clear whether one can choose, or bound, the cardinality of the finite-valued semantics to reduce the maximum satisfiability problem in the infinite valued semantics to the finite valued semantics efficiently, in other words to make the method of finite-valued reductions competitive (to the method of Theorem \ref{thm_maxsat_sat}), our 
benchmark solution in the area. Section \ref{s_tableau} outlines an analytic method of addressing satisfiability and maximum satisfiability in $\standardL$ with preprocessing via a Tseitin transformation, 
using a variant of the approach of \cite{Hahnle:MIP,Olivetti:TableauxInfLuk},
where each branch of a tableau tree ends with solving a system of linear constraints.
The method is proved sound and complete (in the sense explained therein).

\section{Preliminaries and problem formulations}
\label{s_formulation}

The language of propositional {\L}ukasiewicz logic $\logic{\mathrmL}$, denoted $\lang{L}(\Luk)$, 
has two basic connectives:  $\neg$ (negation, unary) and $\oplus$ (strong disjunction, 
binary).
Other connectives are definable as follows:\\
$1$ is $x\oplus\neg x$ and $0$ is $\neg 1$;\\
$x\odot y$ is $\neg(\neg x\oplus\neg y)$ (strong conjunction);\\ 
$x\to y$ is $\neg x \oplus y$ (implication);\\
$x\vee y$ is $(x\to y) \to y$  (weak disjunction) and $x\wedge y$ is $\neg (\neg x \vee \neg y)$ (weak conjunction).\\
Formulas containing these connectives can be expanded into formulas in the basic language with a polynomial increase in size (see, e.g., the Remark in \cite[p.~20]{Aguzzoli-Ciabattoni:Finiteness}), 
% easy proof by induction
so we will freely use the defined connectives.
We do not include the connective $\leftrightarrow$ (equivalence) among the defined connectives, and whenever
$x\leftrightarrow y$ appears, we read it as $(x\to y)\odot(y\to x)$. We  will not need nested
equivalences.

Well-formed formulas $\Fm$ of $\lang{L}(\Luk)$ are built up from 
a denumerable set of propositional variables $\Vbl=\{ x_i \}_{i\in N}$ using 
the connectives of $\lang{L}(\Luk)$.

\begin{definition} 
\label{d_notation}
For any set $M$, let $|M|$ denote the cardinality of $M$. Furthermore, if $\varphi$ is a formula of $\lang{L}(\Luk)$ and if $\langle \varphi_1$, \ldots, $\varphi_m \rangle$ is a list of formulas of $\lang{L}(\Luk)$ for some $m\in N$, then 
\begin{itemize}
\item $\sharp(\varphi)$ denotes the number of occurrences of propositional variables in $\varphi$;
\item $\sharp\langle \phi_1,\dots,\phi_m\rangle$ denotes $\Sigma_{i=1}^m \sharp(\varphi_i)$; 
\item for $S = \langle \phi_1,\dots,\phi_m\rangle$, let $\mathbf{L}(S)$ denote the \emph{set} of subformulas of the formulas in $S$.\footnote{The set of course includes each of the formulas $\varphi_1,\dots,\varphi_m$ and 
each variable occurring in any of them.}
\item $l(S)$ denotes the number of pairwise distinct subformulas\footnote{Here $\varphi$ is viewed as a string, a subformula is a substring, and subformulas are the same if and only if the strings are. Thus $x\oplus (x\oplus x)$ is distinct from $(x\oplus x)\oplus x$. Under this convention, it is trivial to decide whether two formulas are the same.}  of the formulas in the list $S$ (i.e, $|\mathbf{L}(S)|$). 

\end{itemize}
\end{definition}

Given that $\neg\neg\alpha \leftrightarrow \alpha$ is a theorem of $\mathrmL$, 
we will assume double negation (i.e., the collocation $\neg\neg$) does not occur in formulas. 
Given a formula in which a double negation occurs, we first remove any such occurrences of double negations, obtaining an equivalent formula.
With this convention in place, 
the number of occurrences of connectives in $\varphi$ is bounded by $c\sharp(\varphi)$ for a suitable integer constant $c$,\footnote{For example, $c=3$ is suitable in this sense.
To see that, consider that a binary tree with $n$ leaves has $n-1$ inner nodes. Each node (no matter whether internal or a leaf), can be expanded with at most one negation, which 
adds no more than $2n-1$ inner nodes, so the total number of inner nodes is no more than $3n-2$.}
so the total number of subformulas in $\varphi$ is bounded by $c'\sharp(\varphi)$ for another suitable integer $c'$.
Thus $\sharp(\varphi)$ is a good enough notion of the length of $\varphi$ for our purposes.

In {\L}ukasiewicz logic, $\oplus$, $\odot$, $\wedge$ and $\vee$ can be shown to be commutative and associative.
In many cases where the inner structure of formulas does not matter, therefore, one may write $x_1 \oplus \dots \oplus x_n$ without worrying about the order and the parentheses in the ``sum''. 
In such cases  we will use $nx$ for $\underbrace{x\oplus \dots \oplus x}_{n \text{ occurrences}}$, and likewise 
 $x^n$ for $\underbrace{x\odot  \dots \odot x}_{n \text{ occurrences}}$.
Notice that this is a notational convention; strictly speaking an expression ``$\phi^n$'' is not a formula, and 
if  $n$ is represented in a positional notation (say base $p)$, the expression ``$\phi^n$''  is substantially shorter than the formula it represents;
namely while $\phi^n$ is of size $O(\log_p n + \sharp (\varphi))$, the formula  $\underbrace{x\odot  \dots \odot x}_{n \text{ occurrences}}$ is of size
$O(n \sharp(\varphi))$.
Moreover $\wedge$ and $\vee$ are idempotent, $\vee$ and $\wedge$ distribute over each other, and $\odot$ 
and $\oplus$ distributes over $\wedge$ and $\vee$.
 
MV-algebras can be introduced using Mundici's $\Gamma$-functor \cite{Chang:ANewProof,Mundici:AbelianLGr}: 
any MV-algebra is isomorphic to $\Gamma(\alg{G},u)$ for a lattice-ordered Abelian group
$\alg{G}$ with a strong unit $u$. 
In particular, define 
\begin{itemize}
\item $\neg a$ as $u-a$, and
\item $a\oplus b$ as $u \land (a+b)$ 
\end{itemize}
for any $a,b\in G$, where on the right-hand side, the symbols $\{+,-,\land\}$ belong to the language of $\alg{G}$.
Then $ \Gamma(\alg{G},u) = \langle [0,u], \oplus, \neg \rangle$ is an MV-algebra, where the interval is taken in the sense of the group ordering.

The standard MV-algebra $\standardL$ is $\Gamma(\alg{R},1)$, that is, one obtains it from the additive group of real numbers $\alg{R}$.
In particular, the semantics of the basic connectives in $[0,1]$ is as follows:
for any assignment $v$, 
\begin{itemize}
\item $v(\neg\varphi) = 1 - v(\varphi)$, and
\item $v(\varphi\oplus\psi) = \min (1, v(\varphi) + v(\psi))$.  
\end{itemize}
Any assignment $v$ to the variables extends to all well-formed formulas in language $\lang{L}(\Luk)$
under the interpretation provided by $\standardL$.
This fixes the meaning of semantic notions such as tautologousness or satisfiability.
For each formula $\varphi$ of the language $\lang{L}(\Luk)$, we can also
denote $f_\varphi$ the function which this formula defines in the algebra $\standardL$.
Recall that McNaughton \cite{McNaughton:FunctionalRep} characterized the class of functions definable by formulas of $\lang{L}(\Luk)$  in $\standardL$ as 
continuous, piecewise linear with integer coefficients. %see for example ?? for details.

Boolean algebras form a subvariety of MV-algebras; in particular, they can be conceived as
algebras in the language $\lang{L}(\Luk)$ (see for example \cite[Section 3.4.5]{Galatos-JKO:ResiduatedLattices}).
Moreover, in every nontrivial MV-algebra $\alg{A}$, the set consisting of
its bottom element $0^{\alg{A}}$ and its top element $1^\alg{A}$ is closed under all operations
of $\alg{A}$ and the subalgebra of $\alg{A}$ on the universe consisting of these two elements is isomorphic to the two-element Boolean algebra. 
In the  Boolean subalgebra, the interpretations of the connectives $\oplus$ and $\vee$ coincide, 
as do the interpretations of $\odot$ and $\wedge$.
In particular, the language $\{\neg,\vee, \odot\}$ is functionally complete in the two-element Boolean subalgebra.

The set of satisfiable formulas in $\standardL$ is 
$$\SAT \coloneqq \{\varphi \in \Tm \mid \exists v (v(\varphi)=1)\}.$$
In words, a formula  $\phi$ of the language $\lang{L}(\Luk)$  is satisfiable provided at least one assignment in $[0,1]$ 
to the variables in $\phi$ yields the value $1$, in the semantics of connectives provided by the algebra $\standardL$.
It is worth stressing here that 
\begin{itemize}
\item unless stated otherwise, $\phi$ is an arbitrary well-formed formula of $\lang{L}(\Luk)$ as introduced above, without restriction on the syntax;\footnote{As opposed to a typical situation in the Boolean case, where it is common to consider only formulas in conjunctive normal form. Having instances given in a conjunctive normal form is a practical concern, given that Boolean SAT solvers typically assume this restriction. See our definition of {\bf Boolean-SAT}
in Section \ref{s_easy_hard} below.}  
 \item in this paper our default semantics is $\standardL$, which is why the problem is named simply $\SAT$; and
 \item our results do not speak to \emph{different} definitions of satisfiability, such as the one requiring instead that at least one assignment send the formula to a value greater than $0$. 
\end{itemize} 

In analogy to the definitions further within this paper, 
one can also introduce the same problem $\SAT$ in a more computational style, via its characteristic function (on the domain of all well-formed formulas of $\lang{L}(\Luk)$): 
\medbreak
\noindent
{\bf SAT}\\
{\bf Instance:} a formula $\phi$ of $\lang{L}(\Luk)$ in variables $x_1,\dots,x_n$.\\
{\bf Output:} {\bf `true'} if there is an assignment $v$ to $x_1,\dots,x_n$ in $\standardL$ that satisfies $\phi$; 
{\bf `false'} otherwise.   
\medbreak
Analogously, we formulate both the optimization and the decision version of the maximum satisfiability problem for the semantics $\standardL$.
\medbreak
\noindent
{\bf MaxSAT-OPT}\\
{\bf Instance:} a multiset $\langle \phi_1,\dots,\phi_m\rangle$ of formulas of $\lang{L}(\Luk)$ in variables $x_1,\dots,x_n$.\footnote{We accept the convention that $x_1,\dots,x_n$ is a list of variables each of which occurs in at least one $\phi_i$ for $1\leq i \leq m$.}\\
{\bf Output:} the maximum integer $k\leq m$ such that there is an assignment $v$ in $\standardL$ to $x_1,\dots,x_n$ that satisfies at least $k$ formulas in  the multiset $ \langle \phi_1,\dots,\phi_m \rangle $.  
\medbreak
\noindent
{\bf MaxSAT-DEC}\\
{\bf Instance:} a multiset $\langle \phi_1,\dots,\phi_m \rangle$ of formulas of $\lang{L}(\Luk)$ in variables  $x_1,\dots,x_n$, and a positive integer $k\leq m$.\\
{\bf Output:} {\bf `true'} if {\bf MaxSAT-OPT}$(\langle \phi_1,\dots,\phi_m \rangle)$ is at least $k$, {\bf `false'} otherwise.

\medbreak

In the standard MV-algebra $\standardL$, for any fixed $k\geq 1$, the set of rational numbers $\{\frac{i}{k}\}_{i=0}^k$ in the domain $[0,1]$ is closed under all operations of $\standardL$ and thus, when equipped with the restrictions of the these operations, it forms a subalgebra.
This is a generalization of  what has been said above concerning the bottom and the top element of the domain $[0,1]$. 
We will denote such algebras $\finiteMV{k}$ if the denominator of the fractions is $k$, so in particular  for $k=1$ the algebra $\finiteMV{1}$ is term-equivalent to the two-element Boolean algebra.

It is straightforward to adapt the above definitions of the satisfiability and maximum satisfiability problems to the semantics given by the subalgebra on the $k+1$-element domain, for a given $k$.
We will accordingly write  {\bf SAT}$_k$, {\bf MaxSAT-OPT}$_k$, {\bf MaxSAT-DEC}$_k$ for these variants of the problem
 (i.e., the denominator of the domain is $k$). In the case $k=1$, i.e., for the subalgebra on the domain that includes just the bottom and the top of the interval $[0,1]$, we are dealing with the Boolean semantics, so  {\bf SAT}$_1$ stands for satisfiability of an \emph{arbitrary} formula of $\lang{L}(\Luk)$ in the two-element Boolean algebra. Below we will introduce Boolean satisfiability and Boolean maximum satisfiability
computational problems as they are usually considered, namely for instances that are multisets of clauses, representing a conjunctive normal form.

Throughout we shall use $N$ to denote the set of natural numbers (nonnegative integers) and $R$ for the real numbers, where $+$, $-$, $\leq$, are operation and relation symbols defined in the real numbers usual.
Let ${\mathbf A}$ be an integer $m \times n$ matrix. 
Let ${\mathbf x}$ be an $n$-vector of variables and let ${\mathbf b}$ be an
integer $m$-vector. The {\bf solvability of the system 
of inequalities} ${\mathbf A}{\mathbf x} \leq {\bf b}$ in $R$ can be tested in polynomial time \cite{Schrijver:TheoryLinearIntegralProgramming}.

More generally, for the system ${\mathbf A}{\mathbf x} \leq {\bf b}$, 
 one can ask about the maximal size (i.e., maximal number of lines) of a subsystem that is solvable in $R$. 
This problem is known as the \emph{maximum feasible subsystem} \cite[Section 1]{AmaldiKann:ComplexApproxMaxFS} of a system of linear constraints:
the solution is a natural number $k$ between $0$ and $m$ (with $m$ the total number of lines in the system).
This problem is $\NP$-hard. We shall refer to this optimization problem as {\bf Max-FS}. Notice that the system is not defined as a set, so the same  constraint may appear
multiple times.

There are many variants of the {\bf Max-FS} problem, see e.g.~several suggestions in \cite{AmaldiKann:ComplexApproxMaxFS}.
We will use a variant that partitions the linear constraints into two groups: those that need to be satisfied by any feasible assignment (often called \emph{hard constraints}; the paper \cite{AmaldiKann:ComplexApproxMaxFS} refers to them as ``mandatory'') and those among which we want to satisfy as many as possible, over all feasible assignments (often called \emph{soft constraints}; \cite{AmaldiKann:ComplexApproxMaxFS}  refers to them as ``optional''). Suppose we have a {\bf Max-FS} instance with $m_0$ hard constraints and $m_1$  soft constraints. 
If any feasible assignment to the hard constraints exist, the solution to the problem is between $0$ and $m_1$; if not, then the solution can be set to be a distinct value (such as $-1$). 
This variant of Max-FS problem will be called {\bf Max-FS with hard and soft constraints}.

\section{Some formulas easy for {\bf SAT} and hard for {\bf MaxSAT}}
\label{s_easy_hard}

\begin{definition}{\rm \cite[Def.~7.1]{Hanikova-Savicky:SAT}} 
\label{def_boolean_coonectives_inluk}
Consider the full language \lang{L}$(\Luk)$, including the definable connectives.
\begin{itemize}
\item[(i)] A \emph{literal} is a variable (such as $x$) or a negation thereof (such as $\neg x$);
\item[(ii)] a \emph{$\vee$-clause} is any $\vee$-combination of literals;
\item[(iii)] a \emph{$(\odot, \vee)$-formula} is built up from literals using an arbitrary finite combination of the connectives $\odot$ and $\vee$.
\end{itemize}
\end{definition}

\begin{lemma}{\rm (\cite[Lemma 7.2 and the proof of Theorem 7.4]{Hanikova-Savicky:SAT}) }{\ }
\label{lm:convex}
\begin{itemize}
\item[(i)] Let $\phi$ be a $(\odot, \vee)$-formula with $n$ variables. Then $f_\phi$ is a convex function in $[0,1]^n$.
\item[(ii)] Any $(\odot, \vee)$-formula (in particular, any $\vee$-clause) is satisfiable in $\standardL$ if and only if it is satisfiable in the two-element Boolean algebra $\{0,1\}$.
\end{itemize}
\end{lemma}

Given that $(\odot, \vee)$-formulas define convex functions, they do not capture all McNaughton functions. On the other hand, any Boolean function can be represented by an $(\odot, \vee)$-formula, so this syntactic fragment of the set of well-formed formulas of $\lang{L}(\Luk)$
 is functionally complete with respect for the Boolean semantics. 

We now give a slight generalization of a lemma from \cite{Hanikova-Manya-Vidal:MaxSAT2023}, with a simpler proof. 

\begin{lemma}{\rm \cite[Lemma 2]{Hanikova-Manya-Vidal:MaxSAT2023}} 
\label{lm:convex_formula}
Let $\varphi_1,\dots,\varphi_l$ be a list of $(\odot,\vee)$-formulas in variables $\{x_1,\dots,x_n\}$. 
Assume $\mathbf{a} \in [0,1]^n$ is any vector of reals such that $\varphi_i(\mathbf{a})=1$ for each $1\leq i\leq l$. Then there is a Boolean vector $\mathbf{b}\in \{0,1\}^n$ such that $\varphi_i(\mathbf{b})=1$ for $1\leq i\leq l$.
\end{lemma}

\begin{proof}
Given the list $\varphi_1,\dots,\varphi_l$, let $\gamma$ be the formula $\varphi_1 \odot\dots \odot \varphi_l$.
Then $\gamma$ is an $(\odot,\vee)$-formula and therefore $f_\gamma$ is a convex function on $[0,1]^n$. 
Assume  $\mathbf{a} \in [0,1]^n$ is such that $f_{\varphi_i}(\mathbf{a})=1$ for each $1\leq i\leq l$. 
We have $f_\gamma(\mathbf{a})=1$. 
Towards a contradiction, assume that $f_\gamma(\mathbf{b})=0$ for each $\mathbf{b} \in\{0,1\}^n$.
Then by convexity of $f_\gamma$, we have that $f_\gamma$ on $[0,1]^n$ is a constant function with value $0$,
which contradicts the assumption that $f_\gamma(\mathbf{a})$  has the value $1$.
Thus indeed there is a $\mathbf{b} \in \{0,1\}^n$ such that $f_\gamma(\mathbf{b})=1$. 
\end{proof}

Let us recall the definition of Boolean satisfiability and maximum satisfiability. 
As usual, Boolean formulas are considered in a conjunctive normal form, 
and the conjunction is represented as a multiset (i.e., a list) of clauses. 
The Boolean language will be mapped into the language $\lang{L}(\Luk)$
by taking $\neg$ for the Boolean negation, $\vee$ for the Boolean disjunction, and $\odot$ for the Boolean conjunction,
so in particular the Boolean clauses are just the $\vee$-clauses from Definition \ref{def_boolean_coonectives_inluk} above;
we say just ``clauses''.

\medbreak

\noindent
{\bf Boolean-SAT}\\
{\bf Instance:} multiset $\langle C_1,\dots,C_m\rangle$ of  clauses in variables 
$x_1,\dots,x_n$.\\
{\bf Output:} {\bf `true'} if there is a Boolean assignment satisfying all clauses; {\bf `false'} otherwise.   

\medskip

\noindent
{\bf Boolean-MaxSAT-OPT}\\
{\bf Instance:} multiset $\langle C_1,\dots,C_m\rangle$ of clauses in variables 
$x_1,\dots,x_n$.\\
{\bf Output:} the maximum integer $k\leq m$ such that there is a Boolean assignment that satisfies at least $k$ clauses in  the multiset $ \langle C_1,\dots, C_m \rangle$.   

\medbreak\noindent
The decision version---{\bf Boolean-MaxSAT-DEC}---is defined in the obvious way.

\begin{theorem}
\label{maxsat_for_fragment_is_boolean}
Let $C_1,\dots C_m$ be a list of clauses in variables $x_1,\dots,x_n$
and let $0\leq k\leq m$ be an integer. Then 
{\bf Boolean-MaxSAT-OPT}$(\langle C_1,\dots, C_m \rangle) \geq k$ if and only if  
{\bf MaxSAT-OPT}$(\langle C_1,\dots, C_m \rangle) \geq k$.  
\end{theorem}

\begin{proof}  Let  $k\in N$ be fixed. 
Assume first that  {\bf Boolean-MaxSAT-OPT}$(\langle C_1,\dots, C_m \rangle) \geq k$; that is, there is a Boolean assignment $v$ to $x_1,\dots,x_n$ which sends at least $k$ clauses in the list $(\langle C_1,\dots, C_m \rangle)$ to $1$. Since the two-element Boolean algebra
is isomorphic to a subalgebra of $\standardL$, $v$ also satisfies at least $k$ clauses in $\standardL$ and so 
{\bf MaxSAT-OPT}$(\langle C_1,\dots, C_m \rangle) \geq k$. 
Conversely, assume that {\bf MaxSAT-OPT}$(\langle C_1,\dots, C_m \rangle) \geq k$.  
Then there is an assignment $v'$ in $\standardL$ that sends at least $k$ clauses in the list to $1$. Let $S$ be the list of those clauses
from among $\langle C_1,\dots, C_m \rangle$ that receive the value $1$ under $v'$; by assumption $|S|\geq k$. 
By Lemma \ref{lm:convex_formula}, the clauses in $S$ are also simultaneously satisfied by some Boolean assignment. 
Therefore {\bf Boolean-MaxSAT-OPT}$(\langle C_1,\dots, C_m \rangle) \geq k$.
 \end{proof}

We are ready to showcase the type of formulas of \lang{L}$(\Luk)$ for which satisfiability is easy to decide
but maximum satisfiability is hard. The results above pave the way of relying directly on analogous results 
in the Boolean case. It is well known that Boolean satisfiability for lists of clauses with one or two literals per clause can be decided in polynomial time.

\begin{corollary}
For instances that are $\odot$-conjunctions of clauses with one or two literals per clause,
{\bf SAT} is in $\Po$.
\end{corollary}

\begin{theorem}
For instances that are multisets of clauses with one or two literals per clause,
{\bf MaxSAT-DEC} is $\NP$-hard.
\end{theorem}

\begin{proof}
By \cite[Theorem 1]{Garey-Johnson-Stockmeyer:SomeSimplifiedNPC},
{\bf Boolean-MaxSAT-DEC} for instances where clauses have one or two literals per clause is $\NP$-hard. 
Theorem \ref{maxsat_for_fragment_is_boolean} tells us that in the special case when inputs are  lists of clauses, 
the problem {\bf Boolean-MaxSAT-DEC} coincides with the problem {\bf MaxSAT-DEC}. 
This formally gives a (Karp) reduction of {\bf Boolean-MaxSAT-DEC} to {\bf MaxSAT-DEC} for instances where clauses have one or two literals each, hence the latter problem is $\NP$-hard. 
 \end{proof}

Summing up this section, we have taken the Boolean conjunction, disjunction, and negation on the elements $0$ and $1$ of the domain
 of $\standardL$ to be rendered as $\odot$, $\vee$, and $\neg$ respectively, and then the satisfiability problem for CNF formulas in the Boolean semantics coincides with the satisfiability problem in the semantics given by $\standardL$. 
We have explored some consequences of this fact, namely the sharp divide between the complexity of 2-SAT and 2-MaxSAT problems in the $\standardL$ semantics 
(subject to $\Po\neq\NP$). 

These results can be contrasted with those obtained via a different rendering of the Boolean connectives, which was considered in \cite{BofillManyaVidalVillaret:NewComplexityResults}. The \emph{simple clausal form} considered in that paper is a $\wedge$-conjunction of $\oplus$-disjunctions of literals in the $\standardL$-semantics. Lemma 2 from 
\cite{BofillManyaVidalVillaret:NewComplexityResults} states that satisfiability (in the $\standardL$-semantics) 
for formulas in simple clausal form is decidable in linear time. 
It is interesting to notice that formulas in simple clausal form are once again functionally complete for the Boolean semantics,
as in the case of $(\odot,\vee)$-formulas. 
But clearly formulas in simple clausal form are far removed from being a faithful representation of Boolean satisfiability. 
For example, consider the formula $2x \wedge 2(-x)$, where $x$ is any variable; it is not satisfiable by any Boolean assignment, 
however it is satisfied by the value $\nicefrac{1}{2}$. 
Reasoning dually to Lemma \ref{lm:convex}, if $\phi$ is a formula in simple clausal form, then $f_\phi$ is a  concave function on its domain.
Therefore, formulas in simple clausal form that are unsatisfiable by Boolean assignments can still be satisfied by arbitrary assignments from $\standardL$.

\section{Canonical method}
\label{s_canonical}

At the start of this section, we provide a polynomial-time, many-one (i.e., Karp) reduction of {\bf MaxSAT-DEC} to $\SAT$.
In words, we show how to solve the decision version of the maximum satisfiability problem by posing a single query to an algorithm which solves the satisfiability problem, and returning the answer (where in either case, satisfiability is meant in the semantics is given by the algebra $\standardL$).
Our reduction is similar to those used in \cite{PretoManyaFinger:linkingLukMaxSat}
(which, in turn, refers to  \cite{Mundici:UlamGameMaxSATandMVPart}) and  in \cite{Hanikova:ComplexityValDegree}. 
The differences arise from the fact that, in our case, an unsatisfied formula can take any value below $1$ (not necessarily $0$), and this needs to be addressed in the definition of the set of formulas in the reduction.

Let $\langle \phi_1,\dots,\phi_m \rangle(x_1,\dots,x_n)$ and $k\leq m$ be an instance of the {\bf MaxSAT-DEC} problem.
It is well known that one can implicitly define any rational value in $\standardL$ 
with a formula of $\lang{L}(\Luk)$:
an early example of suitable formulas can be found in \cite{Torrens:CyclicElements}.\footnote{See also \cite[Lemma 3.3.11]{Hajek:1998} for a variant,
and \cite{Hanikova:ImplicitDefConstLuk} for a polynomial-size defining formula, at the cost of increasing the number of variables.}
Let $k\geq 2$ be an integer and let $y$ be a new variable, not among $(x_1,\dots,x_n)$, and let
$$ \rho_{1/k} \coloneqq  \ \ y \leftrightarrow \neg ( (k-1) y)$$

Then we have that $\rho_{1/k}$ \emph{implicitly defines} the rational value $1/k$ in 
$\standardL$ (see, e.g., \cite[Lemma 2]{PretoManyaFinger:linkingLukMaxSat}): 
that is, an assignment $v$ in $\standardL$ sends
$\rho_{1/k}$ to $1$ if and only if it sends $y$ to $1/k$. 
The subformula $ (k-1) y$ of $\rho_{1/k}$ features $k-1$ occurrences of the variable $y$;
notice that $k \leq m$, so the length of $\rho_{1/k}$ is linear in the size of the instance.

For $1\leq i \leq m$, consider a new variable $y_{i,k}$, and let 
$\Phi_{\phi_i,k}$ be the set of formulas
$$ \{ \mbox{ } (\phi_i \leftrightarrow k\, y_{i,k}) \vee \neg y_{i,k}  \mbox{\ \ ,\ \ }   (y_{i,k} \leftrightarrow y )\vee \neg y_{i,k} \mbox{ } \}.$$
Again the length of the formulas in $\Phi_{\phi_i,k}$ is easily seen to be linear in the size of the instance.
Finally, let $\Phi_{k}$ be the list of formulas $\bigcup_{1\leq i\leq m} \{\Phi_{\phi_i,k}\}$.

\begin{theorem}
\label{thm_maxsat_sat}
The pair $\langle \phi_1,\dots,\phi_m \rangle(x_1,\dots,x_n)$ and $k$ with  $2\leq k\leq m$ 
belongs to {\bf MaxSAT-DEC} if and only if the set
  $\{ \rho_{1/k} \} \cup \Phi_{k} \cup \{ \bigoplus_{i=1}^m y_{i,k}\}$  belongs to $\SAT$.
\end{theorem}

\begin{proof}
For the left-to-right direction, assume $v$ to be an assignment satisfying---without loss of generality---the first $k$ formulas of the list. Consider then the assignment $v'$ that coincides with $v$ on the variables $x_1, \ldots, x_n$ and puts $v'(y) = 1/k$ and 
\[v'(y_{i,k}) = \begin{cases} 1/k&\hbox{ if } i \leq k\\ 0 &\hbox{otherwise.}\end{cases}\]
The assignment $v'$ clearly satisfies $\rho_{1/k}$. Next, since $v'(y_{1,k}) = \ldots = v'(y_{k,k}) = 1/k$, also $v'(\bigoplus_{i=1}^m y_{i,k}) = 1$. Lastly, the formulas in $\Phi_{k}$ are satisfied under $v'$: the formulas $(y_{i,k} \leftrightarrow y )\vee \neg y_{i,k}$ are trivially satisfied, since each $y_{i,k}$ is indeed sent to either $1/k$ (and hence, $v'(y)$) or to $0$. For the other formulas in $\Phi_{k}$, first  $v'(\varphi_j) = 1$ and $kv'(y_{j,k}) = k1/k = 1$ for each $1 \leq j \leq k$, and $v'(\neg y_{j,k}) = 1$ for $k < j \leq m$, hence they are all satisfied.

For the right-to-left direction, let $v$ be an assignment satisfying $\{ \rho_{1/k} \} \cup \Phi_{k} \cup \{ \bigoplus_{i=1}^m y_{i,k}\}$. From $\Phi_{k}$ and $\rho_{1/k}$  we know $v(y_{i,k})$ is either $1/k$ or $0$. Therefore, for 
$v(\bigoplus_{i=1}^m y_{i,k}) = 1$, necessarily at least $k$ many $y$-variables are evaluated to $1/k$. Assume, again without loss of generality, that $v(y_{1,k}) = \ldots = v(y_{k,k}) = 1/k$. From $\Phi_{k}$, we get that $v((\phi_i \leftrightarrow k\, y_{i,k}) \vee \neg y_{i,k}) = 1$ for each $1\leq i \leq m$. In particular, since $v(\neg y_{j,k}) \neq 1$ for every $1 \leq j \leq k$, necessarily $v((\phi_j \leftrightarrow k\, y_{j,k}))$ for each such $j$. Together with the previously observed fact that $y_{j,k} = 1/k$ for each such $j$, this implies that  $v(\phi_1)= \ldots = v(\phi_k) = 1$, concluding the proof.
\end{proof}

\noindent
For $k=1$, it is immediate that $\langle \phi_1,\dots,\phi_m\rangle$ and $k$ is in {\bf MaxSAT-DEC} 
if and only if $(\dots(\varphi_1\vee\varphi_2)\vee \dots )\vee \varphi_m$ is in   $\SAT$.
Given that for $m=k=1$ both problems coincide, we get:

\begin{corollary} 
\label{cor_maxsat_npc}
The problem {\bf MaxSAT-DEC} is  $\NP$-complete.
\end{corollary}

This reduction from {\bf MaxSAT-DEC} to {\bf SAT} provides a practical approach to the 
MaxSAT problem in $\standardL$, 
provided that we use a competitive algorithm for solving $\SAT$ (i.e., 
the satisfiability problem in $\standardL$). 
We could rely on either of the following two $\SAT$ solvers, which have been shown rather efficient. 
The first one is the tableau with constraints 
method proposed by H\"ahnle \cite{Hahnle:MIP} that reduces $\SAT$ to Mixed Integer Programming (MIP) 
and can therefore use any available MIP solver. 
The second one is the Satisfiability Modulo Theory (SMT) methods proposed by Ans\'otegui et al.\ that 
reduces $\SAT$ to an SMT satisfiability problem and can use any available SMT 
solver \cite{AnsoteguiBofillManyaVillaret:SATLukSMT12,AnsoteguiBofillManyaVillaret:SMT16,Vidal:MNiblos}. 
These methods can take advantage of the latest developments and innovations in MIP and SMT solvers, 
avoiding the need to implement a $\SAT$ solver from scratch.

Furthermore, a polynomial-time Turing reduction (a.k.a.~Cook reduction) of {\bf MaxSAT-OPT} to {\bf MaxSAT-DEC} 
can be given, as we proceed to explain.  
It is this approach that prompts our referring to this method of solving {\bf MaxSAT-OPT} 
as \emph{canonical}, given its wide scope of applicability 
in the realm of optimization problems (see, e.g., \cite{Stockmeyer:CC}).
This kind of reduction can be viewed as using an unspecified algorithm for {\bf 
MaxSAT-DEC} as an \emph{oracle}; the difference to a Karp reduction is that any number of calls to the oracle
can be made, and their number and order can be determined on the basis of the oracle answers to earlier queries in the same computation. 
As usual with oracle computations, any call to the oracle
counts as one step in the computation, and under this proviso, 
the oracle computation needs to run in time polynomial in the input size, in our case  $\sharp\langle \phi_1,\dots,\phi_m\rangle$. 
% That is, $\Sigma_{i=1}^m \sharp(\varphi_i).$
Using the terminology and notation of \cite{Papadimitriou:CC}, 
the class of function problems that can be solved in polynomial time with an oracle in the complexity class $\NP$ is denoted $\FPNP$. 
In our case, given an instance $\langle \phi_1,\dots,\phi_m\rangle$, it is easy to arrive at the optimal value for {\bf MaxSAT-OPT} 
using binary search on the discrete, polynomial-size search space $\{1,\dots,m\}$ of possible solutions, using at most $\lceil\log m\rceil$ oracle calls to {\bf MaxSAT-DEC}. 
Now considering that {\bf MaxSAT-DEC} is  $\NP$-complete by Corollary \ref{cor_maxsat_npc}, we have the following:

\begin{corollary}
\label{lm:maxsat-opt-in-fpnp}
{\bf MaxSAT-OPT} is in $\FPNP$.
\end{corollary}

For membership in this complexity class, it is not important that the oracle solves 
 {\bf MaxSAT-DEC}; any oracle solving an $\NP$-complete problem (an $\NP$-oracle) 
would suit, and indeed one can use any algorithm for $\SAT$, combining it with  our 
Theorem \ref{thm_maxsat_sat} given above.
In view of the obvious reduction from {\bf MaxSAT-DEC} to {\bf MaxSAT-OPT},  
the two problems are equivalent in the sense that if 
either has a polynomial-time algorithm, so does the other one.

Cook reductions allow a polynomial number of calls to the oracle but we have used at most $\lceil\log m\rceil$ many.
Can one do better than $O(\log m)$ oracle calls?  
Below, we provide a classification of the problem in terms of Krentel's work \cite{Krentel:ComplexityOptimization} 
that suggests a negative answer subject to $\Po \not= \NP$.
Krentel ranks  optimization problems in $\FPNP$
in terms of the number of calls to an $\NP$-oracle.
For $z: N\longrightarrow N$ a smooth function (i.e., $z$ is 
non-decreasing and polynomial-time computable in unary representation), 
$\FPNP[z(n)]$ is the class of functions computable in polynomial time with an $\NP$ oracle 
with at most $z(\abs{x})$ oracle calls for instance $x$, where $\abs{x}$ denotes the length of $x$.
By definition, $\FPNP$ coincides with $\FPNP[n^{O(1)}]$ since a polynomial-time algorithm can make no more than a polynomial amount of oracle calls.

For $\Sigma$ a finite alphabet let $f,g: \Sigma^\ast \longrightarrow N$.  
A \emph{metric reduction} \cite{Krentel:ComplexityOptimization} 
from $f$ to $g$ is a pair $(h_1,h_2)$ of 
polynomial-time computable functions where $h_1:\Sigma^\ast \longrightarrow \Sigma^\ast$ and 
$h_2:\Sigma^\ast \times N \longrightarrow N$ such that
$f(x)=h_2(x,g(h_1(x)))$ for all $x\in\Sigma^\ast$. 
The notion of a metric reduction is a natural generalization of polynomial-time many-one reduction 
to optimization problems. It follows from the definition that for each smooth function $z$ as above, 
$\FPNP[z(n)]$ is closed under metric reductions. 

\begin{theorem}{\rm (\cite{Krentel:ComplexityOptimization}, see also \cite{Stockmeyer:CC})}
Assume $\Po\not=\NP$. \\
Then $\FPNP[O(\log\log n)] \subsetneq \FPNP[O(\log n)] \subsetneq \FPNP[n^{O(1)}]$.
\end{theorem}

Krentel \cite[Theorem 2.2 and discussion on p.~500]{Krentel:ComplexityOptimization} proves that {\bf Boolean-MaxSAT-OPT} is complete for $\FPNP[O(\log m)]$ under metric reductions. This implies the following result.

\begin{theorem}
{\bf MaxSAT-OPT} is complete for $\FPNP[O(\log m)]$ under metric reductions.
\label{th_complete_fpnp}
\end{theorem}

\begin{proof}
Containment was obtained in Corollary \ref{lm:maxsat-opt-in-fpnp}
and the discussion preceding it.

We prove hardness. Our claim is that a metric reduction, as defined earlier in this section following Krentel, of {\bf Boolean-MaxSAT-OPT} to {\bf MaxSAT-OPT} is provided by a pair of identity functions (in the role of the two functions denoted $h_1$ and $h_2$ in the definition).

Take an arbitrary instance of {\bf Boolean-MaxSAT-OPT} problem: a multiset $\langle C_1,\dots,C_m\rangle$ of  clauses in variables $x_1,\dots,x_n$.
Clearly {\bf Boolean-MaxSAT-OPT}($\langle C_1,\dots,C_m\rangle$) is a natural number $k$ where $1\leq k \leq m$.
By Theorem  \ref{maxsat_for_fragment_is_boolean}, the same number $k$ is also the correct answer to 
{\bf MaxSAT-OPT}($\langle C_1,\dots,C_m\rangle$).

This shows that one can correctly solve the function problem {\bf Boolean-MaxSAT-OPT}
by passing each instance (lists of clauses) to any available algorithm for {\bf MaxSAT-OPT}, solving it there, and reporting back the output.
\end{proof}

The binary search algorithm always makes a logarithmic number of oracle calls, 
no matter what the instance is.
Also, the complexity analysis as given does not take into account the efficiency of the computations
executed by the oracle; all that we know about the oracle is that it correctly decides a particular 
$\NP$-complete problem.
Considering the experience obtained in Boolean MaxSAT solvers based on Boolean SAT solvers, 
there might be alternatives to binary search that might turn out to be more efficient in practice,
where one departs from the paradigm that emphasizes worst-case complexity. 
A typical Boolean MaxSAT solver does a \emph{linear} search, either from 
unsatisfiable to satisfiable (core-guided approach), or from satisfiable to unsatisfiable (model-guided approach) \cite{bacchus2021maxsat,LiManya:2021maxsat}.
The solvers heavily exploit the fact that the formulas in the multiset are Boolean clauses (i.e., a \emph{normal form} is assumed) and that a SAT solver also returns a satisfying assignment or an unsatisfiable core; moreover, the calls to the SAT solver need not be its independent runs. These parallels invite an openness of mind when implementing
MaxSAT solvers for the MV-semantics.

\section{Analysis of finite-valued semantics}

\label{sec:redToClassic}

This section proposes a direct solution for the maximum satisfiability problems over a finite subalgebra of $\standardL$. In what follows, we let $N$ be a fixed natural number, constant throughout the section, and we will talk about {\bf MaxSAT-DEC}$_N$ and {\bf MaxSAT-OPT}$_N$,  
 restricting assignments to the algebra $\MVN \coloneqq \{0, \frac{1}{N}, \ldots, \frac{N-1}{N},1\}.$ 
After setting up a few technical lemmas, 
 we will give a reduction of {\bf MaxSAT-DEC}$_N$ to {\bf Boolean-SAT}  (Lemma \ref{lem:DEC-VALMAXSATtoCLASSAT}) and of 
 {\bf MaxSAT-OPT}$_N$  to {\bf Boolean-Partial-MaxSAT} (Lemma \ref{lem:VALMAXSATtoCLASPARTMAXSAT}).
At the end of the section we will discuss to which degree, in combination with existing results in the literature,
these results could be taken as an alternative method to solve {\bf MaxSAT-OPT} and \textbf{DEC} (infinitely valued), 
with the target problems being  {\bf Boolean-SAT} and {\bf Boolean-Partial-MaxSAT}, respectively. 

Consider an instance
$S \coloneqq \langle \varphi_1, \dots, \varphi_m\rangle(x_1, \ldots, x_n)$ of {\bf MaxSAT-OPT} (or the corresponding fragment of an instance of {\bf MaxSAT-DEC}), 

and recall (from Definition \ref{d_notation}) that ${\mathbf L}(S)$ denotes 
the \emph{set} of all subformulas of the formulas in the list $S$, and for the complexity considerations,  $l(S) \coloneqq \vert {\mathbf L}(S)\vert$ (the cardinality of ${\mathbf L}(S)$).

To encode the information about ${\mathbf L}(S)$ and the possible values they can take in the $(N+1)$-valued MV-algebra, we introduce the set of 
variables
\[V\coloneqq \{x_{\chi, j}\mid \chi \in {\mathbf L}(S), j \in \{0, \ldots, N\}\}\]
Each variable $x_{\chi, j}$ is intended for a Boolean semantics, where it takes the value  $1$ if and only if, in the fixed  $(N+1)$-valued MV algebra $MV_N$,  the formula $\chi$ takes the value $\frac{j}{N}$.

In order to simplify the notation, for $i,j \in \{0, \ldots, N\}$ we write $j \oplus i \coloneqq \min\{N, j+i\}$.
To capture the behaviour of the MV-operations (under a given assignment) we will use the following set $HC$ (the set of hard constraints):

\begin{align*}
(\mathrm{1a})& \bigvee_{0 \leq i \leq N} x_{\varphi, i}, &\text{ for each }\varphi \in {\mathbf L}(S), \\
(\mathrm{1b})&\, \neg x_{\varphi, j} \vee \neg x_{\varphi, i} , &\text{ for each }\varphi \in {\mathbf L}(S) \text{ and }0 \leq i < j \leq N, \\
(2)&\, \neg x_{\varphi, j} \vee x_{\neg \varphi, N-j}, & \text{ for each }\neg \varphi\in {\mathbf L}(S) \text{ and } 0 \leq j \leq N,\\
(3)&\,  \neg x_{\varphi, j} \vee \neg x_{\psi, i} \vee x_{\varphi \oplus \psi, a} & \text{ for each } \varphi \oplus \psi \in {\mathbf L}(S), 0 \leq a \leq N,\\
 & &\text{ and }{i \oplus j = a\text{ with  }0 \leq i,j\leq N}
\end{align*}
Thus $HC$ is a list of clauses, intended to be considered for simultaneous satisfaction (in the Boolean semantics).
Then, as we shall see, formulas in (1a) and (1b) guarantee that among the set of variables $\{x_{\varphi, j}\}$ with $0 \leq j \leq N$, a single one will be true (since, indeed, $\varphi$ takes a single value in $MV_N$). Formulas in $(2)$ encode the behavior of the negation, and those in $(3)$, that of the strong disjunction. 
We do not treat the definable connectives here, for simplicity.

\begin{observation}
The size of $HC$ (i.e., $\sharp HC$) is polynomial in the size of $S$ (i.e, $\sharp S$). 
\end{observation}
\begin{proof} The total number of occurrences of subformulas in $S$ is linear in  $\sharp S = \Sigma_{i=1}^m \sharp(\varphi_i)$ (see the discussion under Definition \ref{d_notation}), and $l(S)$ is bounded by that value.  Furthermore, we can estimate the number of literals in $HC$ as the the sum of:
    \begin{itemize}
        \item[(1a)] $l(S) \times N$ literals, 
        \item[(1b)] at most $2 \times l(S) \times N^2$ literals, 
        \item[(2)] at most $2 \times l(S) \times N$ literals,   % for each value, its negation 
        \item[(3)] at most $3 \times l(S) \times N^2$ literals. % for each two values, their Lukasiewicz sum
    \end{itemize}
We remark that $N$ is fixed and does not depend on the input.
\end{proof}

\begin{lemma}\label{prop:HCboolsat}
Let $e$ be a Boolean assignment satisfying all formulas in $HC$. Then, for each $\varphi \in {\mathbf L}(S)$, there is a unique integer $1 \leq i_\varphi \leq N$ such that $e(x_{\varphi_i, i_\varphi}) = 1$. Moreover, the assignment defined by $e^*(p) \coloneqq \frac{i_p}{N}$ for each propositional variable $p$ occurring in $S$ satisfies, for each $\varphi\in {\mathbf L}(S)$,
\[e^*(\varphi) = \frac{i_\varphi}{N}\]
and is therefore a sound assignment in the finite MV-algebra in the algebra $\MVN$. 
\end{lemma}
\begin{proof}
The first claim is immediate since $e$ satisfies formulas in $(1)$. We will prove the second claim by induction on the complexity of the formula $\varphi$. 

The base case, namely the variables occurring in $S$, is given in the definition. 

Let $\varphi = \neg \psi \in {\mathbf L}(S)$. By induction hypothesis, $e^*(\psi) = \frac{i_{\psi}}{N}$, which means by definition of $i_{\psi}$ that  $e(\neg x_{\psi, i_{\psi}}) = 0$. Since $e$ satisfies the clauses in $(2)$, we get that $e(x_{\neg \psi, N-i_{\psi}}) = 1$, meaning that $i_{\neg \psi} = N-i_{\psi}$. By definition of the MV-negation, $e^*(\neg \psi) = 1-e^*(\psi) = \frac{N-i_{\psi}}{N}$, which concludes the proof of the step.

Let $\varphi = \psi \oplus \chi \in {\mathbf L}(S)$. By i.~h., $e^*(\psi) = \frac{i_{\psi}}{N}$ and $e^*(\chi) = \frac{i_{\chi}}{N}$, so by definition, 
$e(\neg x_{\psi, i_{\psi}}) = 0$ and $e(\neg x_{\psi, i_{\chi}}) = 0$. Then, from the formulas in $(3)$, we get that $e(x_{\psi \oplus \chi, i_\psi \oplus i_\chi}) = 1$, meaning that $i_{\psi \oplus \chi} = i_{\psi} \oplus i_{\chi}$. On the other hand, by definition of the strong MV-disjunction $\oplus$, $e(\psi \oplus \chi) = \min\{1, e(\psi) + e(\chi)\} = min\{1, \frac{i_{\psi} + i_{\chi}}{N}\} = \frac{i_{\psi} \oplus i_{\chi}}{N}$ (last equality by definition of $i \oplus j$), concluding the proof.
    
\end{proof}

\begin{lemma}\label{prop:valEvtoBoolEv}
Let $h$ be an assignment in $\MVN$ to the variables occurring in $S$. Then the Boolean assignment $h'(x_{\varphi,j}) \coloneqq \begin{cases}  1&\hbox{ if }h(\varphi) = \frac{j}{N},\\ 0 &\hbox { otherwise} \end{cases}$ for each formula $\phi$ in ${\mathbf L}(S)$ satisfies all formulas in $HC$. 
\end{lemma}
\begin{proof}
It is proven by induction on the complexity of the formula in $HC$ under consideration.

Formulas in (1) are trivially satisfied, since for each $\psi$ there is a unique $0 \leq j \leq N$ for which $h(\psi) = \frac{j}{N}$. 

For formulas in (2), consider the formula $\neg \psi$ and assume that $h(\psi) = \frac{j}{N}$, so $h'(\neg x_{\psi, j}) = 0$. Then, $h(\neg \varphi) = 1-h(\varphi) = \frac{N-j}{N}$, by definition of the MV-negation, meaning that $h'(x_{\neg \psi, N-j}) = 1$.

For formulas in (3), consider the formula $\varphi \oplus \chi$. We know that $h(\varphi) \oplus h(\chi) = h(\varphi \oplus \chi)$. Therefore, for $i,j$ such that $h(\varphi) = \frac{j}{N}, h(\chi) = \frac{i}{N}$, and hence, such that $h'(\neg x_{\varphi, j}) = 0$, $h'(\neg x_{\chi, i}) = 0$), it holds by definition that $h(\varphi \oplus \chi) = \frac{i \oplus j}{N}$, and so, that $h'(x_{\varphi \oplus \chi, i \oplus j}) = 1$.
\end{proof}

\begin{lemma}\label{lem:DEC-VALMAXSATtoCLASSAT}
There is a polynomial-time many-one reduction of the problem  {\bf MaxSAT-DEC}$_N$ to the problem {\bf Boolean-SAT}. 
\end{lemma}
\begin{proof}
First we refer to a Boolean cardinality constraint that expresses that at least $k$ among $m$ Boolean variables are set to $1$. 
We apply the standard polynomial-size sequential counter introduced in \cite{Sinz:towardsOptimalCNF}, which encodes the ``at most'' constraint, 
from which the ``at least'' is obtained by taking negated Boolean variables (relying on the equivalence $\leq (m-k) (\neg y_1,\dots, \neg y_m) \iff \geq k ( y_1,\dots, y_m)$, spelled out in the notation of \cite{Sinz:towardsOptimalCNF}). Let $C(y_1,\dots,y_m, z_1,\dots, z_j)$ be such a sequential counter, namely a set of Boolean clauses, polynomial size in $m$, and expressing  $\geq k ( y_1,\dots, y_m)$; the variables $z_1$ to $z_j$ are the encoding variables.

We claim that the pair $ \langle \phi_1, \dots, \phi_m\rangle(x_1, \ldots, x_n)$ and $k$ belongs to  {\bf MaxSAT-DEC} if
and only if $HC \cup \{C(x_{\phi_1,N},...,x_{\phi_m,N},z_1,\dots,z_j)\}$ belongs to {\bf Boolean-SAT}.
Indeed suppose at least $k$ among $\langle \phi_1, \dots, \phi_m\rangle$ are satisfied by some assignment $v$ in $\MVN$;
w.l.o.g. let the first $k$ formulas be true under $v$. Then from  Lemma \ref{prop:valEvtoBoolEv} we obtain a Boolean assignment $v'$ satisfying $HC$ and in particular, we have $v'(x_{\phi_i,N}) =1$ for $1\leq i\leq k$, therefore $v(C)=1$.
On the other hand, if $w$ is a Boolean assignment satisfying $HC\cup\{C\}$, then since $C$ is true, we have 
$w(x_{\phi_i,N})=1$ for at least $k$ values of $i \in [1,\dots,m]$ and using Lemma \ref{prop:HCboolsat} we obtain a correct assignment $w^\ast$ in $\MVN$ where in particular $w^\ast(\phi_i)=\frac{N}{N}$ for at least $k$ values of $i \in [1,\dots,m]$,
whence  $\langle \phi_1, \dots, \phi_m\rangle(x_1, \ldots, x_n)$ and $k$ belongs to  {\bf MaxSAT-DEC}.

\end{proof}

Instead of the decision problems above, we can consider as a variant, a reduction to (Boolean) maximum satisfiability as an optimization problems with hard and soft constraints. 
That is, the function problem {\bf MaxSAT-OPT} can be reduced to the function problem {\bf Boolean-Partial-MaxSAT} as follows.

\begin{lemma}\label{lem:VALMAXSATtoCLASPARTMAXSAT}
Let $S$ be given, as above.
Consider the {\bf Boolean-Partial-MaxSAT} instance $BS$, given by $HC$ as the set of hard constraints, and, as soft constraints, the list 
$\langle x_{\phi_1,N}, \dots, x_{\phi_m,N}\rangle$. Then $k$ is a solution for $BS$ if and only if $k$ is a solution for $S$. 
\end{lemma}
\begin{proof}
Let us first assume that $k$ is a solution for $S$, namely, there is an assignment $h$  into $[0,1]$ satisfying, w.l.o.g, $\varphi_1, \ldots, \varphi_k$. From above, we can assume the homomorphism $h$ is into $MV_N$. By Proposition \ref{prop:valEvtoBoolEv}, we know that $h'$ satisfies $HC$, and furthermore, 
$h'(x_{\varphi_1, N}) = \ldots, h'(x_{\varphi_k, N}) = 1$, and $h'(x_{\varphi_r,N}) = 0$ for all the other $k < r \leq m$. 

Let us now check the other direction, namely, assume $k$ is a solution for $BS$. This means there is a Boolean assignment $e$ that satisfies all hard constrains and satisfies, wlog, $x_{\varphi_1,N}, \ldots, x_{\varphi_k, N}$.
It is first clear that, since $e$ satisfies formulas in $(1)$, it means that, for each $\psi \in {\mathbf L}(S)$ -and in particular, for each variable  appearing in $S$- there is a single $0 \leq i_{\psi} \leq N$ for which $e(x_{\psi, i_\psi}) = 1$. On the other hand, following Proposition \ref{prop:HCboolsat}, we know that $e^*(\varphi_j) = \frac{i_{\varphi_j}}{N} = 1$ for each $1 \leq j \leq k$, concluding the proof of the lemma.

 \end{proof}

The above approach, which reduces {\bf MaxSAT-DEC}$_N$ directly to {\bf Boolean-SAT}, with a variant of reducing {\bf MaxSAT-OPT}$_N$ to {\bf Boolean-Partial-MaxSAT}, is useful for handling maximum satisfiability problems over a \emph{fixed} finite MV-chain. 
This does not entail, however, that it is advantageous to reduce the maximum satisfiability problem
in the infinitely valued semantics $\standardL$ to a finitely-valued semantics $\MVN$ for a ``suitable'' $N$.

In principle this could be done: it is well known that for any given formula $\phi(x_1, \ldots, x_n)$ there exists a natural number $k$ such that, if $\phi$ is satisfiable in $\standardL$, then $\phi$ is also satisfiable in a finite MV-chain whose cardinality is bounded by $k$. Such upper bounds were originally established in \cite{Mundici:Satisfiability} and later improved in \cite{Aguzzoli:asymptoticBoundLuk,Aguzzoli-Ciabattoni:Finiteness}, by showing that $f_\phi$ in $n$ variables takes its minimum/maximum at a vector ${\mathbf a}$ with rational values, namely a vertex of some $n$-dimensional cell of a polyhedral decomposition $C(\phi)$ of $[0,1]^n$ (the domain of $f_\phi$),
and estimating the common denominator of ${\mathbf a}$. These results were formulated for formulas in the basic language.
In particular, for a list $S = \langle \varphi_1, \dots, \varphi_m\rangle(x_1, \ldots, x_n)$ by \cite[p.~19]{Aguzzoli-Ciabattoni:Finiteness}
we know that $S$ (as a conjunction) is satisfiable in $[0,1]$ if and only if it is satisfiable in $MV_j$ for some $j \leq N_{S}$, where $N_{S} = \frac{(\sharp S)}{n}^n + 1 <  2^{(\sharp S)}$.
Thus their results show that the denominators are polynomial-size when represented in a positional notation.

However in order to obtain a single finite MV-chain  $MV_j$ for which testing the maximum satisfiability problem in  $MV_j$ is guaranteed to yield the same result as in the infinite-valued semantics on $\standardL$, the approach is to define $j$ as
$\mathrm{LCM}\{i \colon i \leq N_S\}$, a quantity that grows exponentially in $N_{S}$ (relying on estimates of the second Chebyshev function).
So this method of finite-valued reduction hardwires an exponential dependence on the input size into the algorithm, 
even for special cases of the decision problem  {\bf SAT}  defined for the infinite-valued semantics, 
where it suffices to find a single cardinality for which the test succeeds. (One might consider  a \emph{nondeterministic} reduction of the maximum satisfiability problem to a classical satisfiability problem, thereby leaving some room for heuristics; in this setting, one would proceed by letting $N$ be a nondeterministically chosen positive integer bounded by $N_{S}$.)

On top of that, in the case of maximum satisfiability, we are not aware of any method that avoids searching through substantially many, possibly all possible cardinalities of the finite chains. This approach performs poorly because it loses any heuristics underlying the reductions from  {\bf SAT} to {\bf Boolean-SAT}, as it forces the explicit computation of all possible values in the largest possible algebra. Therefore, in spite of the fact that finite-valued reductions are theoretically interesting and engendered the seminal complexity results
in {\L}ukasiewicz logic referred to above, unless the cardinality of a finite-valued algebra is fixed in advance, we favour the method presented in Theorem \ref{thm_maxsat_sat}.

\section{Tableau-like method for maximum satisfiability}
\label{s_tableau}

This section provides a tableau approach to solving  {\bf SAT} as well as {\bf MaxSAT-OPT}; recall that both problems were introduce in Section \ref{s_formulation}. We start with the  {\bf SAT}  problem in Subsection \ref{ss_sat_tab} and present the method for {\bf MaxSAT-OPT} as a modification in Subsection \ref{ss_tab_maxsat}. The variant for $\SAT$ is conceptually easier, and at the same time it provides a lower bound on computing 
 {\bf MaxSAT-OPT}, as will become clear. 

\subsection{Satisfiability}
\label{ss_sat_tab}

Our method for  the $\SAT$ problem combines several approaches that might be termed \emph{analytic}. 
$\SAT$ and its complexity have been investigated in depth 
\cite{Mundici:Satisfiability,FingerPreto25,Aguzzoli-Ciabattoni:Finiteness,MundiciOlivetti:Resolution,Aguzzoli:asymptoticBoundLuk,Hanikova:Handbook,SchockaertJanssenVermeir:SATasCSP,Hanikova:Handbook,AnsoteguiBofillManyaVillaret:SATLukSMT12,AnsoteguiBofillManyaVillaret:SMT16,Hanikova-Savicky:SAT,BofillManyaVidalVillaret:NewComplexityResults}. 
In particular, tableau calculi have  been proposed in \cite{Hahnle:MIP,Olivetti:TableauxInfLuk}.
Our method, in its variant generating a tree with an exponential number of branches, 
provides a simple proof for $\SAT$ in $\NP$ and an upper bound on the runtime 
of a deterministic algorithm for $\SAT$. We also present a heuristic which avoids
processing repeatedly the same subformula of the input formula (in case of repetitive occurrences of subformulas).

The method operates in two subsequent stages. The first one is a variant of Tseitin transformation of an arbitrary formula to a formula in a suitable \emph{normal form} \cite{Tseitin:ComplexityDerivations}. In the context of classical logic, usually by  ``normal form'' one understands a conjunctive normal form. In our case, the suitable normal form is a system of equations in the language \lang{L}$(\Luk)$, such that one side of the equation is a variable. This \emph{Tseitin normal form} is obtained by listing subformulas and introducing a new variable for each of them. This transformation preserves satisfiability of the formula, incurs a polynomial increase in size, and adds new variables. A variant of the transformation was used for testing $\SAT$ in \cite[Section 4]{BofillManyaVidalVillaret:NewComplexityResults}.

Recall that if $\phi$ is a formula, we use $l(\phi)$ for the number of pairwise distinct subformulas of $\phi$ (and analogously for lists of formulas, cf.~Definition \ref{d_notation}).
We can illustrate the advantage of considering only pairwise distinct subformulas with the following example.
\begin{example}
\label{ex:3}
Recall the formula $\rho_{1/k}$ from Section \ref{s_canonical}, and its subformula $(k-1)y$. In that section,
it did not matter how brackets were placed in the expression 
$$\kappa \coloneqq (k-1)y = \underbrace{y\oplus y\oplus \dots \oplus y}_{(k-1) \text{ times}};$$
However the placing of brackets does matter in determining $l(\kappa)$.
If brackets in $\kappa$ nest to the right, then $l(\kappa)$ is proportional to $(k-1)$. 
If $(k-1)y$ is bracketed as a balanced (insofar as possible) binary tree, then  $l(\kappa)$ 
is proportional to $\log_2(k-1)$.
\end{example}

The second stage is a tableau-like procedure that utilizes the system of equations obtained in the first stage 
as labels for nodes in a rooted linear tree, and expands the nodes using simple rules that translate these equations of $\lang{L}(\Luk)$
into linear equations in the reals. 
Subsequently, each branch is evaluated for solvability in the reals, analogously to \cite{Hahnle:MIP,Olivetti:TableauxInfLuk}.

The algorithm for $\SAT$ is given below. The presentation is informal. For simplicity, we will assume the formula is given in the basic language $\{\neg,\oplus\}$ of {\L}ukasiewicz logic. If defined connectives are present, we expand them into basic connectives.\footnote{Another option would be to keep the defined connectives and then, in step \ref{tree_step_5}, rewrite them into the language of linear inequalities in $\alg{R}$ in a way analogous to the basic connectives.}

\noindent
\line(1,0){150}\\
\noindent {\bf Decision method} $\mathrm{T}\mathrmL_{\mathrm{SAT}}$.
 Let $\varphi(x_1,\dots,x_n)$ be an input formula.

\begin{enumerate}
\item {\bf List subformulas.} \label{tree_step_1} Let ${\mathbf L}\coloneqq  {\mathbf L}(\phi)$ be the list of all pairwise distinct subformulas occurring in $\varphi$, 
including $\varphi$ and all its variables.
Let $l\coloneqq l(\phi)$ be the number of items in ${\mathbf L}$.
 
W.l.o.g.~we assume that if $\alpha$ is a subformula of $\beta$, then $\alpha$ occurs before $\beta$ in ${\mathbf L}$.
\item {\bf Name subformulas.} \label{tree_step_2} Introduce new pairwise distinct variables $z_i$ for the $i$-th formula in ${\mathbf L}$ with $1\leq i\leq l$.
These will be called ``$z$-variables''.
It is assumed that the $z$ variables are also distinct from each $x_j$ for $1\leq j\leq n$.\footnote{This is 
a convention in favour of clarity of presentation. Avoiding introduction of new variables for atoms 
$x_1,\dots,x_n$ would save $n$ new variables.}
\item {\bf Equations on names.} \label{tree_step_3} Let ${\mathbf S}$ be the list of equations 
in the language $\{\neg,\oplus\}$  obtained by initializing  ${\mathbf S}$  as empty and taking the following step 
for each item in the list ${\mathbf L}$:
\begin{itemize}
\item if $x$ is a propositional variable in $\varphi$ and $1\leq i\leq l$ and $z_i$ is the variable for $x$, include in ${\mathbf S}$ the equation
\begin{center} 
$x=z_i;$
\end{center} 
\item if $\neg\alpha$ is a subformula of $\varphi$  and $1\leq i,j\leq l$ and $z_i$ is the variable for $\alpha$ and $z_j$ is the variable for $\neg\alpha$,
 include in ${\mathbf S}$ the equation 
\begin{center}
 $z_j = \neg z_i;$   
\end{center}
\item if $\alpha\oplus\beta$ is a subformula of $\varphi$ and $1\leq i,j,k\leq l$ and 
$z_i$, $z_j$, $z_k$  are the variables for $\alpha$, $\beta$, $\alpha\oplus\beta$ respectively, 
include in ${\mathbf S}$ the equation 
\begin{center}
$z_i \oplus z_j = z_k.$    
\end{center} 
\end{itemize}
Having each item of ${\mathbf L}$ processed, ${\mathbf S}$ contains equations in the basic language $\{\neg,\oplus\}$
of {\L}ukasiewicz logic. The number of equations in ${\mathbf S}$ is $l$.  
\item {\bf Initialize tree.}
Initialize a rooted tree ${\mathbf T}$, linear at this stage, with $l$ nodes. 
From the root down, 
label each node of ${\mathbf T}$ with one equations from ${\mathbf S}$.
Start with equations containing the $x$-variables and mark them \emph{final}.
Then process those containing $\neg$ and subsequently those containing $\oplus$
and mark each as \emph{active}.\footnote{The structure of ${\mathbf T}$ will be linear up to a certain point and binary from there on. This is the case because a) the equations with the $x$-variables are not expanded, and b) all the equations with $\neg$ are expanded before any of the equations with $\oplus$, and the expansion rule for $\neg$ does not lead to branching.
Cf.~Example \ref{ex:1}.}
\item  
{\bf Boundary constraints.} 
Append before the root $2l$ new nodes labelled $0\leq z_i\leq 1$ for each $i=1,\dots,l$.
Mark each as \emph{final}.
\item {\bf Target constraint.} Append as new root of the tree a node labelled $z_l=1$ 
for $z_l$ the variable introduced for $\varphi$. (By convention taken in step \ref{tree_step_1}, 
$z_l$ is assigned to $\varphi$.) Mark \emph{final}.
\item {\bf Expand tree.} \label{tree_step_5}
 From the root of $\mathbf{T}$ towards the leaves, 
 process each node $N$:
\begin{itemize}
\item If the label of $N$ is marked \emph{final} (i.e, does not contain  $\neg$ or $\oplus$), 
leave it intact and proceed to the next node.
\item If the label of $N$ is marked \emph{active} (contains $\neg$ or $\oplus$), 
mark it \emph{passive}, and below each leaf of ${\mathbf T}$, append a new subtree with labelled nodes using the following {\bf expansion rules} (one new node per each constraint), marking each new label \emph{final}:
\begin{center}
\begin{tabular}{cc}

\begin{tabular}{c|c}  
\multicolumn{2}{c}{$z_i\oplus z_j = z_k$} \\
\hline $z_i + z_j \leq 1 $ &  $z_i + z_j \geq 1$ \\
 $z_i + z_j = z_k $ & $z_k=1$ 
\end{tabular}
& \qquad
\begin{tabular}{c}  
{$z_i = \neg z_j $} \\
\hline $z_i = 1 - z_j$ \\
{\ } 
\end{tabular}

\end{tabular}
\end{center}
\end{itemize}
An application of the rule on the left involves branching below each leaf of ${\mathbf T}$.
The labels in the conclusions of these rules are linear constraints 
in real numbers. The mark \emph{final} indicates the algorithm leaves them intact.
Having processed all nodes of ${\mathbf T}$, each branch of ${\mathbf T}$ defines
a system of linear constraints marked \emph{final} in an unambiguous way. 
\item {\bf Solve systems.} \label{tree_step_8} From the leftmost branch to the right, 
test the system of constraints labelled \emph{final} on the branch for solvability in $\alg{R}$\footnote{The testing procedure is in $\Po$. For the purpose of testing, one can render
each equality ${\mathbf a}{\mathbf x}={\mathbf b}$ as two inequalities ${\mathbf a}{\mathbf x}\leq{\mathbf b}$ 
and $-{\mathbf a}{\mathbf x}\leq-{\mathbf b}$.} 
until a branch is found whose system of constraints is solvable. 
In such a case, return {\bf `yes'} and exit.
\item {\bf Default.} Return {\bf `no'} and exit.
\end{enumerate}

\vskip -15pt
\noindent
\line(1,0){155}

Typically in an analytic tableau method (cf., e.g., ~H\"ahnle's \cite{Hahnle:MIP}), 
one starts with a given formula $\varphi$ and decomposes it, 
taking one occurrence of a connective in each step and
expanding the tableau using the given tableau rules. 
If a subformula occurs multiple times in $\varphi$,
it is processed multiple times and each time, new variables are introduced with it: cf.~\cite[section 5.1]{Hahnle:MIP} where
new variables $i_1$ and $i_2$ are introduced for each occurrence of an implication. 
This is usually considered a feature of the analytic method.
With creating the set of subformulas first, we avoid this and have potentially less new 
variables, see Example \ref{ex:3}. Admittedly, on this score, our method might therefore not qualify as purely analytic and/or internal, 
cf.~the introduction of \cite{Olivetti:TableauxInfLuk} or \cite[Section 4]{Aguzzoli-Ciabattoni:Finiteness}.

\begin{example}\label{ex:1}
A simple example will illustrate the generation of the tree and the resulting systems of constraints.
Consider the formula $((x \oplus \neg y) \oplus \neg (x \oplus y)) \oplus \neg (x \oplus y)$. A list of its subformulas is the following:
\[\langle 
x,
y,
\neg y,
 x \oplus y,
  x \oplus \neg y,
   \neg (x \oplus y), 
    (x \oplus \neg y) \oplus (\neg (x \oplus y)), 
     ((x \oplus \neg y) \oplus \neg (x \oplus y)) \oplus \neg (x \oplus y)
     \rangle \]
In order to present the example in a compact way, we write three initial nodes only: the first, with the boundary, target and ground equations; the second, with the equations from $\mathbf{S}$ 
with symbol $\neg$, and the third, with the equations from $\mathbf{S}$ with symbol $\oplus$. 
Below this, we expand the tree as described by the algorithm. We omit marks (active, passive, final).
We use vertical dots to indicate the tree that would be included in their place is a copy of the one depicted at its side.\\ 

\begin{tabular}[b]{c}

\begin{tabular}[t]{c}

\branchheight{.3in}

\synttree[${\displaystyle z_8 = 1, \{0 \leq z_i \leq 1\}_{1 \leq i \leq 8}, z_1 = x, z_2 = y}$ [${\displaystyle z_6 = \neg z_4, z_3 = \neg z_2,}$ [${\displaystyle z_8 = z_7 \oplus z_6, z_7 = z_5 \oplus z_6, z_5 = z_1 \oplus z_3,  z_4 = z_1 \oplus z_2 }$ [${\displaystyle z_6 = 1 - z_4, z_3 = 1-z_2}$ [${\scriptstyle  z_8 = z_7 + z_6 }$ [${\scriptstyle z_7 + z_6 \leq 1}$ [$\vdots$]]] [${\displaystyle  z_8 =1 }$ [${\displaystyle z_7 + z_6 \geq 1}$  [$ {\displaystyle z_7 = z_5 + z_6 }$ [${\displaystyle z_5 + z_6 \leq 1}$  [$ {\displaystyle z_5 = z_1 + z_3 }$ [${\displaystyle z_1 + z_3 \leq 1}$  [${\displaystyle  z_4 = z_1 + z_2} $ [${\displaystyle z_1 + z_2 \leq 1}$]] [$ {\displaystyle z_4 =1 }$ [${\displaystyle z_1 + z_2 \geq 1}$ ]]]] [${\displaystyle  z_5 =1 }$   [${\displaystyle z_1 + z_3 \geq 1}$  [${\displaystyle  z_4 = z_1 + z_2 }$ [${\displaystyle z_1 + z_2 \leq 1}$ ]] [${\displaystyle  z_4 =1 }$ [${\displaystyle z_1 + z_2 \geq 1}$ ]]]]]] [ ${\scriptstyle  z_7 =1 }$ [${\scriptstyle z_5 + z_6 \geq 1}$  [$ \vdots$]]]]]]]]]
\end{tabular}

\end{tabular}

\end{example}

\begin{lemma}
\label{lm:invariant}
The expansion rules in step \ref{tree_step_5} of\/ $\mathrm{T}\mathrmL_{\mathrm{SAT}}$
preserve the following invariant: for any assignment $v$ in the domain $[0,1]$
to all $z$-variables, $v$ satisfies the equation in the premise in the algebra
$\standardL$ if and only if $v$ satisfies all constraints in at least one branch
in the conclusions of the rule in the algebra $\alg{R}$.
\end{lemma}

\begin{proof} % \ref{lm:invariant}
Notice that the expansion rules work as a switch between the signature of $\lang{L}(\Luk)$ 
and language of $\alg{R}$. 
In both cases the statement is a straightforward consequence of the semantics
of the connectives $\neg$ and $\oplus$ in $\standardL$. 
We prove the case for $\oplus$.
Top-to-bottom: let $v$ be an assignment of values in $[0,1]$ to $z$-variables 
introduced in step \ref{tree_step_2}, and consider $z_i,z_j,z_k$ s.t.~  
$v(z_i)\oplus v(z_j) = v(z_k)$ is true in $\standardL$. Then it must be the case
that either $v(z_i) + v(z_j) \leq 1$ and $v(z_i) + v(z_j) = v(z_k)$ holds in $\alg{R}$, or 
 $v(z_i) + v(z_j) \geq 1$ in which case we also have $v(z_k)=1$ in $\alg{R}$.
 Bottom to top: again  let $v$ be an assignment of values in $[0,1]$ to $z$-variables.
 If $v(z_i) + v(z_j) \leq 1$ and $v(z_i) + v(z_j) = v(z_k)$  both hold in $\alg{R}$,
 we have $v(z_i)\oplus v(z_j) = v(z_k)$ is true in $\standardL$.
 If $v(z_i) + v(z_j) \geq 1$ and $v(z_k) =1 $  in $\alg{R}$,
 we have $v(z_i)\oplus v(z_j) = v(z_k)$ is true in $\standardL$.
 This exhausts possible cases.
\end{proof}

\begin{theorem}
\label{thm:sat_complete}
The  method $\mathrm{T}\mathrmL_{\mathrm{SAT}}$ is sound and complete for $\SAT$.
\end{theorem}

\begin{proof}
The {\bf soundness} claim states that whenever the method answers `yes' on input $\varphi$, 
then there is an assignment $v$ to $x_1,\dots,x_n$ such that $v(\varphi)=1$. 
So assume that there is a branch $B$ of ${\mathbf T}$ such that the system of constraints 
given by $B$ is solvable, under some assignment $v$ to variables on $B$, and fix $v$.
In particular, for $i=1,\dots,n$, the variable $x_i$ gets value $v(x_i)$ (notice each $x_i$
occurs on every branch).
The assignment $v$ extends to $\varphi$ in a unique way and
one shows by induction on the structure of $\varphi$, using Lemma \ref{lm:invariant},
 that for any subformula $\psi$ of $\varphi$, we have $v(\psi) = v(z_j)$ for $z_j$ 
 with $j\in\{1,\dots,l(\phi)\}$
 being the $z$-variable assigned to $\psi$ in step \ref{tree_step_2}. 
In particular, $v(\varphi) =1 $.

The {\bf completeness} claim states that if $v(\varphi) =1$ for some assignment $v$, 
then the method yields `yes' on input $\varphi$.
So fix $v$ s.t.~$v(\varphi) =1$. We claim there is a branch of ${\mathbf T}$ 
with a solvable system of equations.
First produce the full tree ${\mathbf T}$. 
Then  assign values to all $z$-variables,
starting from those that are names for $x_1,\dots,x_n$, 
and then inductively on the structure of $\varphi$
using again that $v(\psi) = v(z_j)$ for a $z_j$ assigned to $\psi$ in step \ref{tree_step_2}. 
This is consistent with equations obtained in step \ref{tree_step_3}. 
By abuse of language, call this assignment $v$.
The assignment $v$ makes it possible to travel downward 
from the root of ${\mathbf T}$ via labelled nodes, using Lemma \ref{lm:invariant}
to show that $v$ satisfies each label: in particular if ${\mathbf T}$ branches due to a
node with label $z_i\oplus z_j = z_k$, then (assuming the label in the premise is satisfied by $v$), 
Lemma \ref{lm:invariant} guarantees
that there is at least one branch on which the new (and hence, all) labels are satisfied by $v$.
Finally a leaf $L$ of ${\mathbf T}$ is reached: since Lemma \ref{lm:invariant} was 
applied at each expansion, and since the boundary and the final constraint clearly hold
under $v$, all final constraints on the branch determined  by $L$ hold under $v$. 
\end{proof}

\begin{lemma}
\label{lm:sizes}
The problem $\SAT$ on an instance $\varphi$ can be solved deterministically 
by constructing the tree ${\mathbf T}$ and testing the solvability 
of systems of linear constraints in $\alg{R}$ on no more than $2^{l(\phi)}$ branches. 
Each branch has at most $4l(\varphi)+1$  constraints and
 $l(\varphi)+n$ variables.
\end{lemma}

\begin{proof}
The tree branches at each occurrence of $\oplus$ in $\mathbf{S}$;
the number of such occurrences is bounded by $l(\phi)$.
Each branch has at most $2l(\phi)$ constraints for subformulas (since
each occurrence of a connective gives rise to at most two constraints) plus $2l(\phi)$ boundary constraints, plus a target constraint.
(Here we do not consider the possibility of replacing each equation with two inequalities.)
Each branch of the tree uses all the variables: $n$ input variables $x_1,\dots, x_n$
and $l(\phi)$ $z$-variables.
\end{proof}

\begin{corollary}
\label{cr:np}
The problem $\SAT$ is in $\NP$, in particular, a formula is satisfiable
if and only if there is a polynomial-size witness consisting of a tableau branch 
of the method  $\mathrm{T}\mathrmL_{\mathrm{SAT}}$ and a matching 
system of constraints solvable in $\alg{R}$.
\end{corollary}

\begin{proof}
Since the method $\mathrm{T}\mathrmL_{\mathrm{SAT}}$ is (sound and) complete for $\SAT$ by 
Theorem \ref{thm:sat_complete}, any satisfiable formula has the following 
polynomial-size certificate of its own satisfiability in $\standardL$:
the system of equations in $z$-variables constructed in step \ref{tree_step_3}, 
and a branch of the tree $\mathbf{T}$, defined by a list of instructions
specifying which branch to take upon each application of $\oplus$-rule, combined with a system $C$ of constraints that matches
the indicated branchings (in the sense that the equations with $\oplus$ have been expanded according
to the specified branch) and such that $C$ is solvable in $\alg{R}$. On the other hand,
the soundness (and completeness) theorem also says that an unsatisfiable formula \emph{cannot} have
such a certificate.
\end{proof}

Furthermore, any decision tree obtained from the above procedure can be linearized, using the methods of mixed integer programming as in \cite{Hahnle:MIP}. In particular, any instance of the application of the branching rule introduced in step \ref{tree_step_5} can be replaced by an instance of an application of the following lemma (observing the condition that distinct Boolean variables will be used for distinct subformulas):
 
\begin{lemma}
\label{lm:linearization}
{\rm cf.~(\cite[sct.~5.1]{Hahnle:MIP}, \cite[Lemma 6.2.19]{Hajek:1998})}
Assume $a_1,a_2,a_3\in [0,1]$. Then $a_1 \oplus a_2 = a_3$ holds in $\standardL$ if and only if
there is an $y\in \{0,1\}$ such that all of the following constraints hold in $\alg{R}$:
\begin{multicols}{2}
\begin{itemize}
\item[(i)] $a_1 + a_2 \leq 1+y$
\item[(ii)] $y\leq a_1 + a_2$
\item[(iii)] $a_3 \leq a_1 + a_2$
\item[(iv)] $a_1 + a_2 \leq a_3 + y$
\item[(v)] $y\leq a_3$.
\end{itemize}
\end{multicols}
\end{lemma}

\begin{proof}
Assume $a_1 \oplus a_2 = a_3$ holds in $\standardL$. 
Case 1: $a_1 + a_2 \leq 1$, then from the assumption we have $a_1 + a_2 = a_3$. We set $y\coloneqq 0$. 
The fact that $a_1,a_2,a_3\in [0,1]$ implies (ii) and (v); the remaining constraints in the Lemma follow from $a_1 + a_2 = a_3$.
Case 2: $a_1 + a_2 > 1$. The assumption implies $a_3=1$; we set $y\coloneqq 1$, we get (v).  
The fact that $a_1,a_2,a_3\in [0,1]$ implies
(i) and (iv). From $a_1 + a_2 > 1$ we get (ii) and (iii).\\
Now assume there is an $y\in \{0,1\}$ such that all constraints listed hold in $\alg{R}$.
Case 1: $y=0$. We have (i) $a_1 + a_2 \leq 1$ and (iii,iv) $a_3 \leq a_1 + a_2 \leq a_3$. 
Hence  $a_1 \oplus a_2 = a_3$.
Case 2: $y=1$. We have (v) $1\leq a_3$ and (ii)  $1\leq a_1 + a_2$.  Hence  $a_1 \oplus a_2 = a_3$.
\end{proof}

This modification eventually yields, in step \ref{tree_step_8}, a single instance of the MIP problem, where the integer variables are \emph{Boolean};
see \cite[6.2.18]{Hajek:1998}. Thus we arrive at one of the competitive ways of addressing the $\SAT$ problem.
A major advantage of using a MIP solver is the advanced possibility of heuristics, whereas in the simple version above, the only optimization considered is aborting the computation upon finding a branch with a solvable system. That is: without the linearization, by design the algorithm  $\mathrm{T}\mathrmL_{\mathrm{SAT}}$ may need to generate and  eventually test for solvability exponentially many systems of equations.\footnote{One might optimize by testing immediately on every generated branch and exiting the computation upon finding one with a solvable system. In our exposition though, we prefer to consider the size of the full decision tree.}  However, from the viewpoint of the worst-case deterministic complexity, the MIP method does not differ substantially from testing on a possibly exponential number of branches.

\subsection{Maximum satisfiability}
\label{ss_tab_maxsat}

In this subsection we adapt the previous method, which addresses the {\bf SAT} problem, to the {\bf MaxSAT-OPT} problem.  It is easily observed that usual methods for $\SAT$, with the method from the previous subsection among them, are not applicable for {\bf MaxSAT-OPT}, even though they easily adapt to test \emph{joint satisfiability} of a list of formulas. See \cite{LiManyaVidal:TableauxMaxSATLuk} for a discussion of this issue.
One problem is of course the differences in both the type of input and the type of output: {\bf SAT} is a decision problem where any method addressing it takes as input a single formula and yields on output a Boolean value, whereas {\bf MaxSAT-OPT} takes a list of formulas and the output is a nonnegative integer. 
The tableau approaches of \cite{Hahnle:MIP,Olivetti:TableauxInfLuk} use MIP solvers on branches, also returning a Boolean value.
Another feature of the method from the previous subsection is that it considers 
distinct subformulas as a set, so any repetition of the same formula in the list on input is obliterated. On the other hand, with {\bf MaxSAT-OPT} we want to take repetitions into consideration, so for example taking any satisfiable formula $\alpha$ and considering the $m$-element list $\langle \alpha,\dots,\alpha\rangle$, for any $m>1$, 
a complete method needs to produce the answer $m$ on this instance.

These considerations invite the approach of preserving the Tseitin-like procedure of listing equations obtained from the subformulas, 
but combining it with: 
\begin{itemize}
\item updating the target constraint  to take into account a multiset of formulas on input, and
\item updating the  query about the system of constraints obtained on each branch, to test maximum feasibility with hard and soft constraints (cf.~Section \ref{s_formulation}).
\end{itemize}
The following algorithm is therefore a modification of the decision method $\mathrm{T}\mathrmL_{\mathrm{SAT}}$ from the previous subsection. 
To highlight the differences, 
each step only gives the information that has changed compared to the previous subsection.

\smallskip\noindent
\line(1,0){150}\\
\noindent {\bf Optimization method} $\mathrm{T}\mathrmL_{\mathrm{MaxSAT}}$ for computing {\bf MaxSAT-OPT}.\\
 Let $\langle \varphi_1,\dots,\varphi_m\rangle$ be a list of formulas in variables $x_1,\dots,x_n$.
 
\begin{enumerate}

\item {\bf List subformulas.}  Let $S \coloneqq \langle \varphi_1,\dots,\varphi_m\rangle$,
let ${\mathbf L}\coloneqq  {\mathbf L}(S)$ 
be the list of all pairwise distinct subformulas occurring in $S$, and 
let $l\coloneqq l(S)$ be the number of items in ${\mathbf L}(S)$.
Conventions as in step \ref{tree_step_1} of $\mathrm{T}\mathrmL_{\mathrm{SAT}}$ apply.
\item {\bf Name subformulas.} As before.
\item {\bf Equations on names.} \label{tree_step_max_3}  As before.
\item {\bf Initialize tree.} As before.

\item {\bf Boundary constraints.} \label{tree_step_max_5}  As before.

\item {\bf Mark hard constraints.}  \label{tree_step_max_6} For each node in ${\mathbf{T}}$ up to this point, 
mark all constraints as {\bf hard constraints}.
\item {\bf Target constraints.} \label{tree_step_max_7} 
Append before the root of  ${\mathbf{T}}$ a new chain 
with labels $z_{j_i} \geq  1$ for $z_{j_i}$ the variable introduced for $\varphi_i$, with $i=1,\dots,m$, preserving the multiplicity of $\varphi_i$ in the input list.  
Mark these constraints as {\bf soft constraints}. 
\item {\bf Expand tree.} \label{tree_step_max_8} As before, preserving in the expansion 
that a hard constraint produces hard constraints. 
\item {\bf Solve systems.}\label{tree_step_10} 
From the leftmost branch to the right, taking one branch at a time.
Each branch defines, via the nodes labelled \emph{final}, 
a system of linear constraints in $\alg{R}$, 
with the target constraints from step \ref{tree_step_5}
marked \emph{soft} and all other constraints marked \emph{hard}.
Thus each branch defines an instance of the Max-FS problem with hard and soft constraints.
Obtain the solution (i.e., an integer) to 
the instance on each branch.\footnote{Since 
all equalities are marked hard, any feasible solution to the {\bf Max-FS} task will
need to satisfy all of them. More generally, see \cite[Concluding remarks]{AmaldiPfetschTrotter:MaxFSIIS} for handling soft constraints
that are equalities.} 
\item {\bf Maximize.} Return the maximum of satisfied soft constraints among  
the constraint systems over all the branches, and exit.
\end{enumerate}

\vskip -15pt
\noindent
\line(1,0){150}

\vskip -5pt

\begin{example}\label{ex:2}
Let us consider the list of formulas  $\langle (x \oplus \neg y) \oplus \neg x, \neg x, x \oplus \neg y, x\rangle$. A list of its subformulas (according to the definition in point 1) is the following:
\[\langle x, y, \neg x, \neg y, x \oplus \neg y, (x \oplus \neg y) \oplus \neg x\rangle \]

In order to depict the example in a compact way we use the same conventions as in Example \ref{ex:1}. Furthermore, we will print in bold the soft constraints.

\begin{tabular}[b]{c}

\begin{tabular}[t]{c}

\branchheight{.3in}

\synttree[${\displaystyle \mathbf{z_6 = 1, z_3 = 1, z_5= 1, z_1 = 1}, \{0 \leq z_i \leq 1\}_{1 \leq 6}}$ [${\displaystyle z_3 = \neg z_1, z_4 = \neg z_2}$ [${\displaystyle z_5 = z_1 \oplus z_4, z_6 = z_5 \oplus z_3}$  [${\displaystyle z_3 = 1-z_1}$ [${\displaystyle z_4 = 1-z_2}$ [${\displaystyle z_5 = z_1 + z_4 }$ [${\displaystyle z_1 + z_4 \leq 1}$ [${\displaystyle z_6 = z_5 + z_3}$ [${\displaystyle z_5 + z_3 \leq 1}$]] [${\displaystyle z_6 = 1}$ [${\displaystyle z_5 + z_3 \geq 1}$]]]] [${\displaystyle z_5 = 1}$ [${\displaystyle z_1 + z_4 \geq 1}$ [${\displaystyle z_6 = z_5 + z_3}$ [${\displaystyle z_5 + z_3 \leq 1}$ ]] [${\displaystyle z_6 = 1}$ [${\displaystyle z_5 + z_3 \geq 1}$]]]]]]]]]

\end{tabular}

\end{tabular}x
\end{example}

\begin{theorem}
The method $\mathrm{T}\mathrmL_{\mathrm{MaxSAT}}$ is sound and complete for {\bf MaxSAT-OPT}.
\end{theorem}

\begin{proof}
 The {\bf soundness} claim states that whenever the method returns $k\in N$ 
on input $\langle \varphi_1,\dots,\varphi_m\rangle$, then there is an assignment $v$ to variables $x_1,\dots,x_n$ that satisfies $k$ formulas
among $\langle \varphi_1,\dots,\varphi_m\rangle$. 
If $\mathrm{T}\mathrmL_{\mathrm{MaxSAT}}$ returns $k$, that means the tree ${\mathbf T}$ was constructed with a branch $B$ and a system of constraints
given by $B$ that yielded $k$ upon solving the Max-FS problem with hard and soft constraints,
and that this was the maximum solution among all branches. Fix such a $v$
and notice that $v$ defines values for $x_1,\dots,x_n$.
Using Lemma \ref{lm:invariant}, all hard constraints from the system, 
 in particular, all constraints 
from steps \ref{tree_step_max_3}, \ref{tree_step_max_5} and \ref{tree_step_max_8} are satisfied by $v$, and so are $k$ of the target constraints.
If $\psi$ is a subformula of some $\varphi_i$ with $i\in\{1,\dots,m\}$, we have $v(\psi) = v(z_j)$ whenever $z_j$ is the $z$-variable assigned to $\psi$, by induction. 
In particular, from step \ref{tree_step_max_7} we have that there are $k$ formulas $\varphi_i$ among $\langle \varphi_1,\dots,\varphi_m\rangle$ such that 
$v(\varphi_i) = 1 $.

The {\bf completeness} claim states that if, for some assignment $v$, there are $k$ items $\varphi_i$ on the list $\langle \varphi_1,\dots,\varphi_m\rangle$ such that $v(\varphi_i) =1$, 
then the method $\mathrm{T}\mathrmL_{\mathrm{MaxSAT}}$ yields at least $k$ on that instance.
So assume that $v(\varphi_i) =1$ for at least $k$ such items and fix $v$. 
We claim there is a branch $B$ of ${\mathbf T}$  with a system of constraints
that yields at least $k$ upon solving its instance of Max-FS problem.
First construct the tree ${\mathbf T}$.
From $v$, we get values for $x_1,\dots,x_n$, the $z$-variables that are their
names, and using equations from step \ref{tree_step_max_3} for the remaining $z$-variables.
The assignment $v$ indicates a leaf of ${\mathbf T}$ that defines a branch $B$ via a series of 
(possibly non-unique) choices on the hard constraints.
If $\psi$ is a subformula of some $\varphi_i$ with $i\in\{1,\dots,m\}$, also 
$v(\psi) = v(z_j)$ whenever $z_j$ is the $z$-variable assigned to $\psi$, all the hard constraints  and at least $k$ soft constraints
are satisfied on $B$ under $v$. Since $k$ formulas on input are satisfied by $v$,
also $k$ soft constraints are satisfied.
Thus the method $\mathrm{T}\mathrmL_{\mathrm{MaxSAT}}$, which returns a maximum over all branches, will yield a value no less than $k$.
\end{proof}

In order to be able to compare the efficiency of the method $\mathrm{T}\mathrmL_{\mathrm{SAT}}$ from subsection \ref{ss_sat_tab} with the method $\mathrm{T}\mathrmL_{\mathrm{MaxSAT}}$ above, putting them side by side, we will assume a small modification of $\mathrm{T}\mathrmL_{\mathrm{SAT}}$ that 
takes as input a finite list of  arbitrary formulas $\langle \varphi_1, \dots,\varphi_m\rangle$ and tests their joint satisfiability. Then both methods take the same type of input, and we obtain comparable trees from both methods, the main difference being in the target constraints.
As we have seen, each branch of the tree obtained from $\mathrm{T}\mathrmL_{\mathrm{SAT}}$ defines a set of linear constraints, the solvability of which is in $\Po$. 
It is typically not necessary to test solvability on all the branches, since the condition under which the method returns  return {\bf `yes'} is existential (namely, this happens if and only if at least one branch defines a solvable system). On the other hand, if  $\langle \varphi_1, \dots,\varphi_m\rangle$ is an input to $\mathrm{T}\mathrmL_{\mathrm{MaxSAT}}$, then on each branch of the generated tree, it is indeed necessary to solve the Max-FS problem with hard and soft constraints
that the branch defines, because the output of the method is defined as the maximum over all the branches. Moreover, the problem on each branch is $\NP$-hard \cite{AmaldiKann:ComplexApproxMaxFS}. In this sense, the complexity of the method $\mathrm{T}\mathrmL_{\mathrm{SAT}}$ is a \emph{lower bound} on the complexity of the method $\mathrm{T}\mathrmL_{\mathrm{MaxSAT}}$. In particular, one can test joint satisfiability of a list $\langle \varphi_1, \dots,\varphi_m\rangle$  using the method 
$\mathrm{T}\mathrmL_{\mathrm{MaxSAT}}$, where the formulas in the list are jointly satisfiable if and only if the output of $\mathrm{T}\mathrmL_{\mathrm{MaxSAT}}$ is $m$.

One can conceive optimizing the method  $\mathrm{T}\mathrmL_{\mathrm{MaxSAT}}$ by observing that, firstly, the multiset of soft constraints remains the same over all the branches, and
secondly, if any subset $S'$ of a set $S$ of hard constraints is unsolvable, then so is $S$.
 We refrain from pursuing these considerations here, since
they are addressed by the methods used in MIP solvers. The following lemma comes in useful.

\begin{lemma}
The tree obtained from the  $\mathrm{T}\mathrmL_{\mathrm{MaxSAT}}$ method can be linearized at the cost of adding
at most $l(\phi)$ Boolean variables. The linearization method does not affect the soft constraints. 
\end{lemma}

\begin{proof}
Any branching in step \ref{tree_step_max_8} of the algorithm can be replaced by expanding the tree with new nodes (without branching)
using Lemma \ref{lm:linearization}. The constraints obtained from the Lemma are all marked \emph{hard}.
This step therefore does not impact the set of possible solutions to the hard constraints in the system. 
The soft constraints are the same on all the branches, therefore the soft constraints in the linearization are well defined.
\end{proof}

The last lemma implicitly extends the definition of the {\bf Max-FS} problem, introduced in Section \ref{s_formulation}, with Boolean variables, in a manner analogous to the case of obtaining the MIP-problem (with Boolean variables) discussed at the end of the previous Subsection. 
This makes it possible to condense an exponential number of {\bf Max-FS} problems into one.
We stress here that the Boolean variables occur exclusively within the hard constraints. We shall refer to such a modification of the {\bf Max-FS} problem as \emph{mixed} {\bf Max-FS} problem. 
Observe that the system of hard conditions % introduced in steps \ref{tree_step_max_6} and \ref{tree_step_max_8} 
in a mixed {\bf Max-FS} problem is always solvable: assign any real values in $[0,1]$ to the variables $x_1,\dots,x_n$, then all the conditions labelled `hard' can be met by computing the values of the $z$-variables (which are new names for the subformulas of the instance). It follows that 
\begin{itemize}
    \item in our case, the solution to the mixed {\bf Max-FS} problem is an integer $k$ with $0\leq k\leq m$, and
    \item the decision version of the mixed {\bf Max-FS} problem is in $\NP$: a positive instance of the problem has a witnessing assignment, consisting of a vector of Boolean values for the Boolean variables in the system, and and a polynomial-size assignment to the real variables showing that at least $k$ of the system of linear conditions obtained by fixing the values of the Boolean variables can be met simultaneously (where $k$ is part of the instance), whereas a negative instance cannot have such an assignment.
\end{itemize}
It is easy to realize that the (optimization version of the) mixed {\bf Max-FS} problem is in $\FPNP$, because it can be solved by a (Cook) reduction to its decision version, with
an asymptotically logarithmic number of queries.
This means that the method introduced in this Subsection is comparable to the one presented in
Section \ref{s_canonical}, where we have provided what we take to be a benchmark solution for {\bf MaxSAT-OPT}, with $\log m$ queries to {\bf MaxSAT-DEC} where the latter is polynomially equivalent to ${\bf SAT}$ (and hence, to MIP).

\section{Concluding remarks}

We have studied a general version of the maximum satisfiability problem in distinguished semantics for {\L}ukasiewicz logic, namely the algebra $\standardL$ and its finite subalgebras. While our work concerns this semantics with an infinite domain, and while we did not consider formulas in a conjunctive normal form, the basic complexity classification we have obtained is analogous to the Boolean case (i.e., the decision version  of the problem is $\NP$-complete in both cases) and there are some other parallels between the two semantics. 

Our work may be taken as exemplifying  good reasons for studying nonclassical logics. While many problems that can be formulated for Boolean semantics also have their version for the MV-semantics, there are nuances of the latter that have no analogy in the former. We have seen an example of this in the discussion of two different ways of rendering the Boolean propositional language into the language of {\L}ukasiewicz logic, at the end of Section \ref{s_easy_hard}, which lead to different properties of the fragment in each case. It is well known (even though not taken into account here) that one can generalize the Boolean notion of a formula being satisfied by an assignment in various ways (e.g., one can require that a formula attain at least a value of a given rational $\nicefrac{c}{d}$, which constitutes the general satisfiability problem, see \cite{MundiciOlivetti:Resolution}, or one can ask, for a given $\phi$, what the maximal value of $f_\phi$ is in $\standardL$, see \cite{Hanikova:ComplexityValDegree}). On a more general scale, nonclassical logics are a fruitful area in which one can define and investigate notions---such as admissible rules---that are trivial in the case of Boolean logic (see \cite{Jerabek:AdmissibleRulesLuk}).

Our work intentionally emphasizes reductions. From the point of view of obtaining lower bounds, this is a necessity, but we also prefer reducing our problems to well-known combinatorial or logical problems for which optimized solvers exist as a method of addressing the upper bounds, rather than attempting a completely standalone solution. As already mentioned, we view this as an advantage, not only providing links to other parts of logic and theoretical computer science, but also from a practical point of view, taking advantage of the heuristics developed there.

\medskip
\noindent
{\bf AI disclosure}
Haniková ran Gemini 3.8 Flash on a mature draft of the manuscript pdf, looking for errors of grammar and style and inconsistencies, as well as to check the accuracy of the paragraph summarizing the contents of the paper. The output was selectively taken into account in improving and clarifying a subsequent version.

\medskip
\noindent
{\bf Funding}
For Haniková and Vidal, this work has been funded by a grant from the Programme Johannes Amos Comenius under the Ministry of Education, Youth and Sports of the Czech Republic, CZ.02.01.01/00/23\_025/0008711.
For Many\`a, this work was supported by grants PID2022-139835NB-C21 and PID2025-171340NB-I00 funded by MCIN/\-AEI/\-10.13039/\-501100011033.

\bibliographystyle{plain}
\bibliography{maxsat}

\end{document}